\documentclass[prb,showpacs,twocolumn,floatfix]{revtex4}
\usepackage{graphicx}
\begin{document}
\def\omegav{{\mbox{\boldmath{$\omega$}}}}
\def\sigmav{{\mbox{\boldmath{$\sigma$}}}}
\def\tauv{{\mbox{\boldmath{$\tau$}}}}
\def\oh{{\scriptsize 1 \over \scriptsize 2}}
\def\of{{\scriptsize 1 \over \scriptsize 4}}
\def\tf{{\scriptsize 3 \over \scriptsize 4}}
\title{Magnetic Structure and Spin Waves in the Kagom\'{e} Jarosite compound
${\bf KFe_3(SO_4)_2(OH)_6}$}

\author{ T.  Yildirim,$^1$ and A. B. Harris,$^2$}

\affiliation{(1) NIST Center for Neutron
Research, National Institute of Standards and Technology,
Gaithersburg, Maryland 20899\\ (2) Department of Physics
and Astronomy, University of Pennsylvania, Philadelphia, PA, 19104}
\date{\today}

\begin{abstract}
We present a detailed study of the magnetic structure and spin waves
in the Fe jarosite compound ${\rm KFe_3(SO_4)_2(OH)_6}$ for
the most general Hamiltonian involving one- and two-spin interactions
which are allowed by symmetry.
We compare the calculated spin-wave spectrum with the recent
neutron scattering data of Matan {\it et al.}
for various model Hamiltonians which include, in addition to isotropic
Heisenberg exchange interactions between nearest ($J_1$)
and next-nearest ($J_2$) neighbors, 
single ion anisotropy and Dzyaloshinskii-Moriya (DM) 
interactions.  We concluded that DM interactions are the dominant 
anisotropic interaction, which not only fits all the splittings in
the spin-wave spectrum but also reproduces the small
canting of the spins out of the Kagom\'e plane.  A brief
discussion of how representation theory restricts the
allowed magnetic structure is also given.
\end{abstract}
\pacs{75.30.Ds,75.10.Jm, 75.50.Ee} 
\maketitle

\section{Introduction}

In search for unusual magnetic ground states and spin dynamics, 
frustrated systems have been the main
focus of both theoretical and experimental investigations 
in recent years.\cite{FSS,ART,sachdev,huse,HKB,mila,sindzingre,
bert,OBD,taner_review} In frustrated magnetic
systems,\cite{FSS,ART} the energies of the various spin interactions
compete and therefore they can not be simultaneously minimized.
This competition can be induced by the geometry of the lattice on which the
spins are arranged, in which case the phenomenon is called
{\it geometrical frustration} and it often leads to
quite unusual low temperature spin structures and dynamics. 
The nearest neighbor (nn) Heisenberg antiferromagnet (AF) 
on the corner-sharing Kagom\'e lattice is one of the most
studied systems since it
has all the ingredients such as low dimensionality, strong frustration, and
low coordination number, required for a disordered "spin liquid" ground
state.\cite{sachdev,huse,HKB,mila,sindzingre,bert}

\begin{figure}
\begin{center}
\includegraphics[width=8cm]{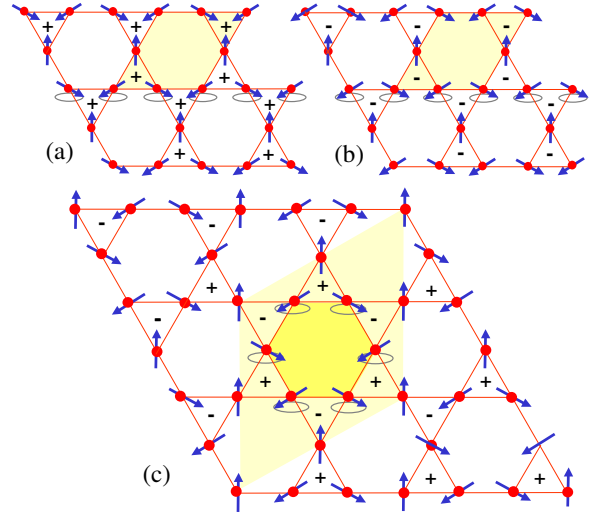} 
\caption{\label{kagome} (Color online) Three possible N\'eel states
of the Kagom\'e lattice nn AF: (a) {\bf q}=0 with positive chirality,
(b) {\bf q}=0 with negative chirality,
and  (c) $\sqrt{3} \times \sqrt{3}$ structure. The plus/minus
signs indicate the chirality of spins on the elemental triangles. 
The shaded (yellow) areas indicate the magnetic unit cells. Note that the
N\'eel sates shown above have continuous degeneracy as the spins
on a line (a-b) and on a hexagon (c) can be rotated out of the 
plane as shown by ellipses without changing the classical ground 
state energy.  }
\end{center}
\end{figure}

Figure~\ref{kagome} shows three possible N\'eel states of the Kagom\'e
lattice nn antiferromagnet. In the classical limit, all ground states
satisfy the ``120$^{\rm o}$ structure", in which the angle between 
each nn pair of spins is 120$^{\rm o}$.
Fig.~\ref{kagome}a-b shows {\bf q}=0 type ordering with respectively
positive and negative chirality. For positive (negative) chirality the 
direction in which the spins rotate as one traverses clockwise
the triangle of sites is clockwise (counterclockwise).
 Fig.~\ref{kagome}c shows the
$\sqrt{3}\times\sqrt{3}$ spin structure which has triangles
with both positive and negative chirality. From the spin configurations shown
in Fig.~\ref{kagome}, it is clear the the classical ground state has a
continuous degeneracy due to the ``weathervane" rotation of the 
spins\cite{sachdev,huse,HKB,mila,sindzingre,bert}
and therefore no long range magnetic order is expected even at zero 
temperature. However, in real systems, there are small perturbations such
as next nearest neighbor (nnn)
interactions,\cite{HKB} anisotropies or defects. It has been also
shown that thermal or quantum fluctuations could also lift some of the
continuous degeneracy known as ``order by disorder," yielding coplanar 
spin structures.\cite{OBD,taner_review} 

\begin{figure}
\begin{center}
\includegraphics[width=5cm]{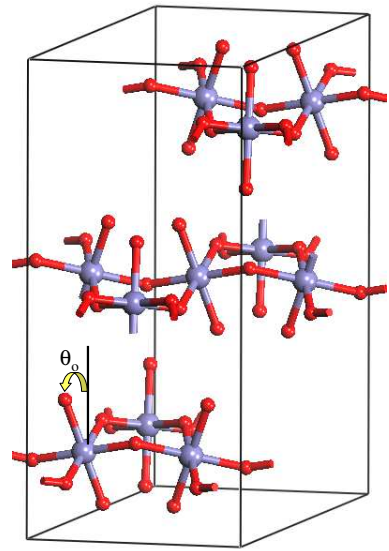} 
\caption{\label{ALUNITE} (Color online) Conventional unit cell of FeJ.
For clarity only Fe, big spheres (blue) and the octahedral oxygens,
small spheres (red) are shown. The tilting  angle, $\theta_o$ of 
the FeO$_{6}$ octahedra is also shown.}
\end{center}
\end{figure}

Despite the extensive theoretical studies that suggest many 
possible fascinating frustrated ground states for the Kagom\'e nn AF, 
few experimental realizations exist for the system.  The initial 
experimental studies were focused on the layered garnet, 
SrCr$_{x}$Ga$_{12-x}$O$_{19}$.\cite{shlee1,keren} 
However the interpretation of 
the magnetic properties of this system were complicated by the
presence of an additional triangular lattice interposed between 
Kagom\'e layers and by the inherent configurational randomness
associated with the random alloying.  In recent years, it has
been shown that the jarosite family of minerals 
${\rm AM_3(OH)_6(SO_4)_2}$,\cite{townsend,
INAMI,WILLS,JAPS,grohol,NAV,barlett,ELHAJAL,YLEE1}
(where A is monovalent ion such as K$^{+}$, and M is trivalent cation 
such as Fe$^{3+}$, Cr$^{3+}$,  or V$^{3+}$) form much better realization
of the two dimensional (2D) Kagom\'e lattice.  As shown in 
Fig.~\ref{ALUNITE} for M=Fe jarosite, the magnetic ion
$M^{3+}$ is centered in a slightly distorted and 
tilted oxygen octahedra, and it forms a Kagom\'e lattice in the 
$ab$-plane. The Kagom\'e planes are widely separated by
nonmagnetic $A^{+}$ and $SO_4^{-2}$ ions. With an exception
of the hydronium jarosite  ${\rm (H_3O)Fe_3(OH)_6(SO_4)_2}$, all members
of jarosite family are found to exhibit long range magnetic
order (LRO) at finite temperatures with varying ground state
spin configurations and exchange interaction strengths depending
on the magnetic ion M. The strength of exchange interaction
is the greatest for the parent M=Fe$^{3+}$ ($d^5, L=0, S=5/2$)
jarosite with Curie-Weiss temperature $\Theta_{CW}=-800 $ K and
$T_N \approx 65 $ K. Replacement of the Fe$^{3+}$ centers by
Cr$^{3+}$ ($d^3, S=1/2)$ also affords an antiferromanetically
ordered material but with a significant reduced 
$\Theta_{CW}=-67 $ K and $T_N \approx 2$ K. Interestingly, when
the magnetic ion is replaced by V$^{3+}$, the ground state
is changed to ferromagnetic Kagom\'e layers which are coupled
antiferromanetically.\cite{NAV}  As we will discuss detail below,
such a ferromagnetic ordering on the Kagom\'e lattice is 
allowed by the  representation theory in contrast to a previous
analysis.\cite{WILLS}

Among the members of the jarosite family, ${\rm KFe_3(OH)_6(SO_4)_2}$
(FeJ) is probably the most studied one.  The magnetic ground state
of FeJ was first investigated by Townsend {\it et al.} using neutron
diffraction.\cite{townsend} However, later it became clear that
the proposed spin configuration in Ref. \onlinecite{townsend} was not
quite correct.\cite{INAMI} Inami {\it et al}\cite{INAMI} presented a
detailed neutron scattering study and determined that the FeJ has 2D
Kagom\'e planes with the {\bf q}=0 spin-structure with positive
chirality as shown in Fig.~\ref{kagome}a. They observed that the unit
cell along the $c$-axis is doubled in the magnetic phase, suggesting the
the Kagom\'e planes are antiferromagnetically coupled. Field-dependent
magnetization measurements\cite{YLEE1}
have suggested that each Kagom\'e plane has
a small ferromagnetic component due to the canting of the spins in
the {\bf q}=0 structure 
with positive chirality (this is the so called "umbrella" configuration).
However to the best of our knowledge, there is no experimental data
which indicates the direction of this spin-canting whose determination
would allow one to deduce the sign of some of the anisotropic terms
in the Hamiltonian, as we will discuss in detail later. 

The fact that FeJ exhibits LRO at finite temperature indicates that 
there are interactions other than nn isotropic AF superexchange,
such as nnn interactions, anisotropies, etc. 
Inami {\it et al.}\cite{INAMI,JAPS} have
pointed out that in FeJ, the oxygen-octahedra are significantly tilted
(see Fig.~\ref{ALUNITE}), thereby inducing a strong single-ion crystal
field (CF) anisotropy. 
They used this CF term in the Hamiltonian to explain the observed
spin configuration and to calculate the spin-wave spectrum.
However in their spin-wave calculations, they neglected the canting
of the spins out of the Kagom\'e plane.  As we will
discuss in detail below, this neglect leads to qualitatively wrong
results for the spin-wave energy gaps at the high symmetry points
in the Brillouin zone.

Elhajal {\it et al.}\cite{ELHAJAL} have pointed out that CF terms
in FeJ should be fairly small because they are second order in the
spin-orbit coupling and Fe$^{3+}$ has a spherical charge distribution
({\it i. e.} $L=0$).  Hence as an alternative source of anisotropy, they
suggested that the Dzyaloshinskii-Moriya (DM) interaction of the
form:\cite{DM1,DM2} $D_{ij} S_i\times S_j$ could stabilize the
experimentally observed spin structure. Indeed, the $z$-component of
the DM vector (the definition of its components is given below)
forces the spins to lie in the $ab-$plane and therefore
effectively acts like an easy-plane anisotropy. The sign of $D_z$
can distinguish between the ${\bf q}$=0 states with negative and
positive chirality. Since the $\sqrt{3}\times \sqrt{3}$ state has
triangles with both positive and negative chirality, the $D_z$ term
alone will not affect the energy of this state.  Furthermore if we
introduce a $D_y$ component, we break the rotational symmetry around
the $c$-axis and create a small anisotropy with respect to
in-plane orientations. The effect of
$D_y$ is also to cant the spins (so that they have a small
out-of-plane component) into the observed
``umbrella" spin configuration. Finally, since the DM interaction
occurs at first-order in the spin-orbit coupling, it is expected to be
larger than the CF terms of Inami {\it et al.}\cite{INAMI,JAPS}

In order to identify the origin of the magnetic interactions that are
responsible for the LRO in FeJ compounds, clearly we need more
experimental data such as the observed spin-wave spectrum. Fortunately,
thanks to the very recent progress in the synthesis of high-quality
single crystals of FeJ compounds,\cite{barlett,YLEE1} such
spin-wave data has been recently become available.\cite{YLEE0}
Here we present a detailed theory for the spin-wave  spectrum in FeJ
compound for generic nn and nnn superexchange interactions including
the CF and DM terms discussed above. As we shall see below, the DM term
along with the nn and nnn isotropic interactions can explain quite well
not only the observed spin configuration but also the observed 
spin-wave spectrum. 

Briefly this paper is organized as follows. In the next section
we first discuss the symmetry of the FeJ structure in the paramagnetic
phase and then present a brief representation analysis of the
allowed magnetic structures. In doing so, we discovered that 
in Wills's analysis,\cite{WILLS} some of the ferromagnetic
wavefunctions were missing.  Indeed such a structure consisting of
ferromagnetic easy-plane Kagom\'e planes
has been observed for a NaV-jarosite.\cite{NAV} 
In this section, we also argue that characterization by
irreducible representations is more fundamental than by the chirality
of the {\bf q}=0 state.  In Sec. III we discuss
the generic magnetic Hamiltonian that we use in our calculations.
In this section, we first discuss the symmetry of the Hamiltonian
matrix and then its representation in a coordinate system in which
the local $z$-axis coincides with the local direction of the spin
moments.  Here we also treat the canting of the spin orientations.
In Sec. IV, we derive analytic results for the spin-wave energy gaps
at the $\Gamma$, $X$ and $Y$-points of the Brillouin zone. In Sec. V,
we present spin-wave
spectra from numerical calculations and compare the results with 
the recent spin-wave data of Matan {\it et al}.\cite{YLEE0} Here we
consider several models with increasing complexity. We find that the
DM terms combined with nn and nnn isotropic exchange interactions
give an excellent fit to the data.  Our conclusions
are summarized in Sec. VI. The details of the diagonalization of the
spin Hamiltonian were given in Appendices. In Appendix A, we
describe the transformation to the boson representations of the
spins.  In Appendix B, we discuss how to obtain the normal mode
energies and eigenvectors. Appendix C describes the actual calculations
of the matrix elements and Appendix D gives the results for the dynamical
structure factor in terms of the normal mode energies and eigenvectors.

\section{Crystal and Magnetic Structure}

\subsection{Symmetry of the Paramagnetic Phase}

\begin{figure}
\begin{center}
\includegraphics[height=5cm]{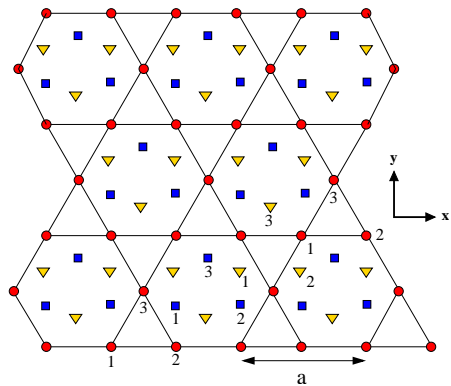} \caption{\label{KPLANE}
(Color online.) Kagom\'e planes.
The filled circles are $x$-$y$ planes at $z=0$,
the filled squares are $x$-$y$ planes at $z=c/3$, and the filled
triangles are $x$-$y$ planes at $z=2c/3$. The positive $z$-axis is
out of the plane. Representative sublattice numbers $\tau$ are given.}
\end{center}
\end{figure}

We first discuss the symmetry of the paramagnetic phase.  In Fig.
\ref{ALUNITE} we show the conventional unit cell which contains
three formula units of FeJ.  The primitive unit cell (which contains
one formula unit of FeJ) is rhombohedral, with basis lattice vectors

\begin{eqnarray}
{\bf a}_1 &=& (a/2) \hat i + (\sqrt 3 a /6) \hat j + (c/3) \hat k \ ,
\nonumber \\
{\bf a}_2 &=& -(a/2) \hat i + (\sqrt 3 a /6) \hat j + (c/3) \hat k \ ,
\nonumber \\
{\bf a}_3 &=& - (\sqrt 3 a /3) \hat j + (c/3) \hat k \ ,
\end{eqnarray}
where $c$ denotes the height of the conventional unit cell of Fig.
\ref{ALUNITE}.  The space group of FeJ is R$\overline 3$m, which
is \#166 in the International Tables of Crystallography.\cite{HAHN}
In order to understand the magnetic properties of FeJ we show in
Fig. \ref{KPLANE} only the Fe $S=5/2$ spins.  Their locations
within the primitive unit cell, denoted $\tauv_n$, are
\begin{eqnarray}
\tauv_1 &=& (0,0,0) \ , \ \ \ 
\tauv_2 = (a/2,0,0) \ , \nonumber \\ 
\tauv_3 &=& (a/4,a \sqrt 3 /4 ,0) \ ,
\end{eqnarray}
where the components are given with respect to the orthogonal
axes of Fig. \ref{KPLANE}.  Notice that each plane
(e. g. the filled circles) forms a Kagom\'e plane lattice.  
As one moves from one plane to the adjacent plane at more
positive $z$, the Kagom\'e lattice is translated
by $(a/2)\hat i + (\sqrt 3 a/6) \hat j$, or, equivalently
by either $(-a/2)\hat i + (\sqrt 3 a/6) \hat j$ or 
$- (a \sqrt 3 /3)\hat j$.

Apart from translations the generators of the space group may be
taken to be ${\cal I}$, spatial inversion about the center of a hexagon,
$r$, a two-fold rotation about an axis parallel to the $x$-axis
and passing through the center of the hexagon, and $R$, a three-fold rotation
about an axis passing through the center of the hexagon perpendicular
to the Kagom\'e plane.  Since the center of a hexagon in one plane
lies just above a triangle in an adjacent plane, the three-fold axis
can be taken at the center of a triangle.  The above operations
imply the existence of mirror planes perpendicular to the Kagom\'e
plane and bisecting the sides of the hexagons.

\begin{figure}
\begin{center}
\includegraphics[height=5cm]{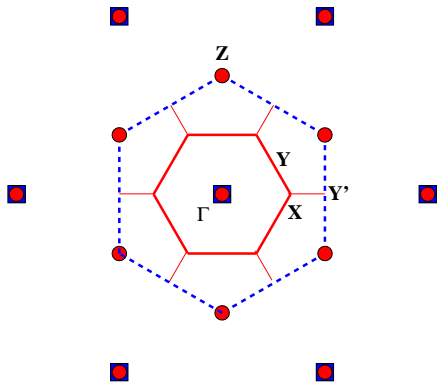} \caption{\label{FEJFIG}
(Color online.) Reciprocal lattices for the FeJ system (blue squares)
and for the two-dimensional hexagonal lattice (red circles).
The first Brillouin zone for the two-dimensional hexagonal system
is indicated by heavy solid lines in red and that (in the $z=0$
plane) for the FeJ system by heavy dashed lines in blue. 
The light solid lines in red mark the boundaries
of neighboring hexagonal zones.  The FeJ Brillouin zone
is three times as large as the hexagonal zone.}
\end{center}
\end{figure}

The reciprocal lattice basis vectors ${\bf b}_i$, which
satisfy ${\bf a}_i \cdot {\bf b}_j = 2 \pi \delta_{i,j}$,
where $\delta_{i,j}$ is the Kronecker delta, are
\begin{eqnarray}
{\bf b}_1/ (2 \pi) &=& (1/a) \hat i + (1/\sqrt 3 a)\hat j + (1/c)\hat k
\nonumber \\
{\bf b}_2/ (2 \pi) &=& - (1/a) \hat i + (1/\sqrt 3 a)\hat j +(1/c) \hat k
\nonumber \\
{\bf b}_3/ (2 \pi) &=& - (2/\sqrt 3 a) \hat j + (1/c) \hat k \ .
\end{eqnarray}
Since interactions between adjacent Kagom\'e planes are very small,
we may approximately describe the structure in terms of a two-dimensional
hexagonal lattice with basis vectors, ${\bf A}_i$, given by
\begin{eqnarray}  
{\bf A}_1 &=& {\bf a}_1 - {\bf a}_2 = a \hat i \ , \nonumber \\
{\bf A}_2 &=& {\bf a}_2 - {\bf a}_3 = -(a/2) \hat i + (\sqrt 3 a/2)
\hat j \ .
\end{eqnarray}
The associated reciprocal lattice of this hexagonal lattice has basis vectors
\begin{eqnarray}
{\bf B}_1/(2 \pi) &=& (1/a) \hat i + (1/\sqrt 3 a) \hat j \ , \nonumber \\
{\bf B}_2/(2 \pi) &=&  (2 /\sqrt 3 a) \hat j \ .
\label{HEXEQ} \end{eqnarray}
In contrast the three-dimensional reciprocal lattice vectors in
the $z=0$ plane for FeJ are
\begin{eqnarray}
{\bf b}_1 - {\bf b}_2 &=& (4 \pi /a) \hat i \nonumber \\
{\bf b}_2 - {\bf b}_3 &=& - (2 \pi /a) \hat i + (2 \pi \sqrt 3/a) \hat j \ .
\label{FEJEQ} \end{eqnarray}
As shown in Fig. \ref{FEJFIG}, the vectors of Eq. (\ref{HEXEQ}) define
the hexagonal Brillouin zones and those of Eq. (\ref{FEJEQ}) define
the boundaries of the FeJ Brillouin zone in the $z=0$ plane.

\noindent
The first Brillouin zone of the FeJ system in the hexagonal plane
has an area three times that of the two-dimensional hexagonal system.
If there were absolutely no interactions between Kagom\'e planes,
the spin-wave dispersion relation in the region in the first
Brillouin zone of the FeJ system but outside that of the\
two-dimensional hexagonal system could be mapped into the
dispersion relation inside the first Brillouin zone of the
two-dimensional system and in this paper we adopt this picture.
For the two dimensional system, spin-wave energies at points
just outside the hexagonal Brillouin zone near point Y are exactly
the same as at the corresponding point just inside the zone near Y.
For the three dimensional FeJ system this is only approximately true.
The spectra of these two points will differ due to interplanar
interactions.  Similarly, for the two-dimensional system the spin-wave
energies at point Z are identical to those at $\Gamma$, whereas for
the FeJ system they will differ due to interplanar interactions.
We will show the spin-wave spectrum observed by Matan et al.\cite{YLEE0}
along the line $\Gamma-X-Y'$, where $Y'$ is equivalent to $Y$
if interplanar interactions are neglected.

\subsection{Results of Representation Theory}

A continuous phase transition may be described by a Landau
expansion in powers of ${\bf S}_\tau({\bf q})$, the Fourier
components of the spin order parameters which are defined as
\begin{eqnarray}
{\bf S}_\tau({\bf q}) &=& (1 \sqrt {N_{\rm uc}}) \sum_{\bf R}
\langle {\bf S}({\bf R}+\tauv ) \rangle e^{i {\bf q}\cdot ({\bf R}+\tauv)} \ ,
\end{eqnarray}
where ${\bf S}({\bf R}+\tauv )$ is the spin at site $\tau$ in the unit cell
at ${\bf R}$, $\langle \ \ \rangle$ denotes a thermal average,
and ${\bf q}$ is the wavevector of the ordering.  At quadratic
order the Landau expansion for the free energy, $F$, assumes the form
\begin{eqnarray}
F &=& \sum_{\tau, \tau'; \alpha , \alpha'} c_{\tau, \alpha ; \tau' ,
\alpha'} ({\bf q})
S_{\tau , \alpha} ({\bf q}) S_{\tau' , \alpha'} (-{\bf q}) \ ,
\label{FEQ2} \end{eqnarray}
where $\alpha$ labels Cartesian components and
$S_{\tau , \alpha} (-{\bf q}) = S_{\tau , \alpha} ({\bf q})^*$.
The ordering that actually occurs corresponds to the eigenvector
associated with that eigenvalue of the quadratic form of
Eq. (\ref{FEQ2}) which first becomes negative (unstable)
as the temperature is lowered starting from the paramagnetic phase.
In view of the invariance of $F$ with respect to the symmetry operations
of the crystal, one can say that the eigenvectors give rise to
an irreducible representation (irrep) of the space group of the
crystal.\cite{GROUP}
This analysis is straightforward and has been given previously,\cite{WILLS}
although the results we present below seem to contradict that reference.

\begin{figure}
\begin{center}
\includegraphics[height=6cm]{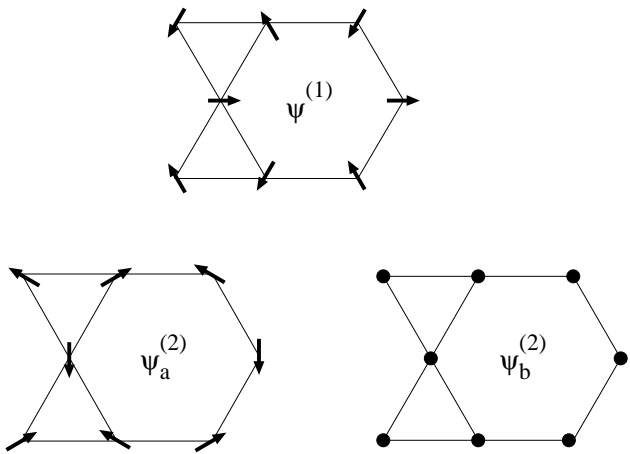} \caption{\label{G1G2}
Spin configurations which transform according to one dimensional
irreps.  Solid circles indicate spins pointing out of the page.
Top: a configuration of $\Gamma_1^+$ for which the eigenvalue
of $r$ is $+1$ and $R$ is $+1$.  Bottom: two configurations for
$\Gamma_2^+$ for which the eigenvalue of $r$ is $-1$ and $R$ is $+1$.}
\end{center}
\end{figure}

Because spin is a pseudovector, spatial inversion takes a spin into
its spatially inverted location but does not reorient the spin.
Since all spins are located at centers of inversion symmetry,
we only need to consider irreps which are invariant
under spatial inversion, and we denote these by
$\Gamma_n^+$, where the label $n$ assumes the values $1,2,3$ and
the superscript indicates invariance under spatial inversion.
For the one dimensional irreps ($\Gamma_1^+$ and $\Gamma_2^+$)
the situation is quite simple: each
eigenvector of $F$ is also an eigenvector of each operation of
the space group.  Thus we show in Fig. \ref{G1G2} the possible spin
configurations which correspond to one dimensional irreps.  The actual
spin configuration associated with irrep $\Gamma_2^+$ is a linear\
combination of the two configurations shown.

\begin{figure}
\begin{center}
\includegraphics[height=8.5cm]{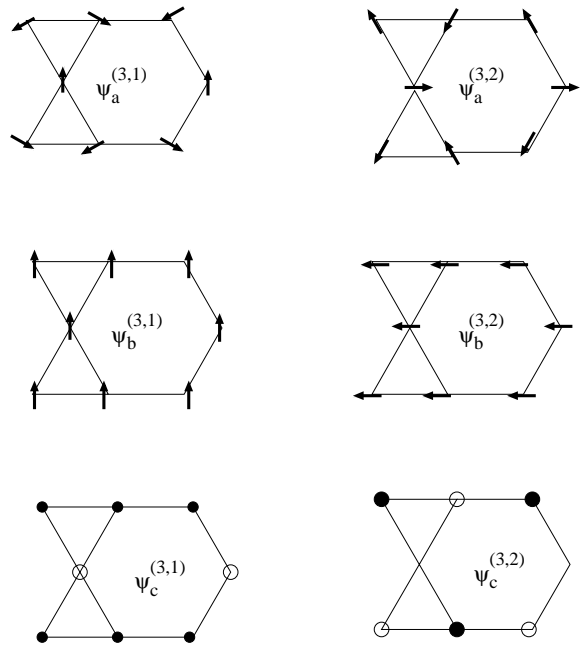} \caption{\label{G3}
Spin configurations which transform according to two dimensional
irrep $\Gamma_3^+$.  In the left (right) column we give the
wavefunctions which transform according to the first (second)
row of the matrices of the irrep. Arrows represent spins of unit length.
Solid (open) circles indicate 
spins pointing perpendicularly into (out of) the Kagom\'e plane.
For $\Psi_c^{(3,1)}$ the magnitudes of the inward spins are
both equal to half that of the outward spin which is set to be
$\sqrt 2$.  For $\Psi_c^{(3,2)}$ the nonzero magnitudes of the spins
are both $\sqrt{3/2}$.} 
\end{center}
\end{figure}

For the two dimensional irrep $\Gamma_3^+$ the situation is
more complicated.  Here one has $p$ wavefunctions $\psi^{(3,1)}_j$,
for $j=1,2, \dots p$ which transform as the first row of
the matrix representation of $\Gamma_3^+$ and their $p$ 
partners $\psi^{(3,2)}_j$ which transform as the second row of
the matrix representation of $\Gamma_3^+$, where for FeJ
$p=3$.\cite{WILLS} We show these wavefunctions in Fig.
\ref{G3}.  To be specific, if ${\cal O}$ is a symmetry operation and
${\cal M}({\cal O})$ is a matrix representation of it according to
$\Gamma_3^+$, then we have

\begin{eqnarray}
{\cal O} \psi^{(3,n)}_j = \sum_m {\cal M}_{n,m}({\cal O}) \psi^{(3,m)}_j \ .
\label{TRANS} \end{eqnarray}
These wavefunctions of $\Gamma_3^+$ correspond to the choice
\begin{eqnarray}
{\cal M}(r) &=& \left[ \begin{array} {c c} -1 & 0 \\ 0 & 1 \\
\end{array} \right] \ , \ \ \ \ 
{\cal M}(R) = \left[ \begin{array} {c c} -1/2 & - \sqrt 3 /2 \\ 
\sqrt 3 /2 & -1/2 \\ \end{array} \right] \ .
\end{eqnarray}
One can check that the wavefunctions shown in Fig. \ref{G3}
actually do satisfy Eq. (\ref{TRANS}).  Of course, all the
wavefunctions shown here are orthogonal to one another and
span the original subspace of spin components.  (That appears not
to be the case with the results of Ref. \onlinecite{WILLS}.
In addition, his states do not seem to allow the net moment for
a single Kagom\'e layer to lie in the Kagom\'e plane.  Since
this does happens for a Na jarosite,\cite{NAV} the statement
that the structure of the chromate jarosite must be the
umbrella structure is not justified.)

\subsection{Discussion of Results: Dipolar Interactions, Chirality}

We now discuss the implications of the above results.  The above
results hold for two separate cases: in the ferromagnetic case,
all Kagom\'e planes have the same spin structure (apart from
a translation), whereas in the antiferromagnetic case successive
Kagom\'e plane have their spins inverted.  The ferromagnetic case
corresponds to wavevector ${\bf q}=0$, whereas the antiferromagnetic
case corresponds to ${\bf q}= (3 \pi /c)\hat k$.\cite{WILLS}
Note that irrep $\Gamma_2^+$ has two wavefunctions.  This means that any
such structure generically consists of a linear combination of a
ferromagnetic state (in a layer) with spins perpendicular to the plane
and some in-plane ordering as well.  This is the so-called ``umbrella''
state.  The implication is that if one has an easy axis ferromagnet
with spins perpendicular to the Kagom\'e plane, one unavoidably
will have some in-plane order.  Conversely, in the case of FeJ
one has order of irrep $\Gamma_3^+$ in each plane which must be
accompanied by a net out-of plane moment.

Note that the irreps $\Gamma_1^+$ and $\Gamma_2^+$ are Ising-like
in that within these representations one can not uniformly rotate
the spins.  Thus the excitation spectrum must have a significant
energy gap.  In contrast irrep $\Gamma_3^+$ is $x$-$y$-like.
One spin structure is some linear combination of the wavefunctions
of the left column of Fig. \ref{G3}.  Alternatively, the free
energy is unchanged if one instead takes the {\it same} linear combination
of the wavefunctions of the right column of Fig. \ref{G3}.  The
actual wavefunction is therefore characterized by four constants,
$\alpha$, $\beta$, $\gamma$ and $\theta$:
\begin{eqnarray}
\psi &=& \cos \theta \left[ \alpha \psi^{(3,1)}_a + \beta \psi^{(3,1)}_b
+ \gamma \psi^{(3,1)}_c \right] \nonumber \\ && \
+ \sin \theta \left[ \alpha \psi^{(3,2)}_a + \beta \psi^{(3,2)}_b
+ \gamma \psi^{(3,2)}_c \right] \ ,
\end{eqnarray}
which can not be fixed by symmetry.  At quadratic level in the Landau
expansion, the free energy does not depend on $\theta$ and one would expect
a Goldstone (gapless) mode.  However, due to the discrete symmetry of the
Kagom\'e plane, the anisotropy energy, $F_A$, which at lowest order is
\begin{eqnarray}
E_A &=& K \cos (6\theta )\ ,
\end{eqnarray}
will lead to a gap.
If $K>0$, the system will favor $\theta=\pi/6$ (or equivalent angles)
whereas if $K<0$ the system will favor angles equivalent to $\theta=0$.
Quadratic energies, such as the dipolar energy, will not lead to a gap. 
Accordingly, it is not surprising that we found that for $\Gamma^+_3$
the dipolar energy of the system was independent of $\theta$.

It is interesting to note that if one assumes isotropic
nearest neighbor (nn) and next-nearest neighbor (nnn) interactions,
then spin configurations $\Psi^{(1)}$ and $\Psi^{(2)}_a$ have
the same energy.  However, in general irrep $\Gamma^{2+}$ will
have lower energy because it can accommodate a distortion into
the configuration $\Psi^{(2)}_b$.  In principle, local single-ion
anisotropy\cite{INAMI} and/or DM interactions\cite{ELHAJAL} can
select a ground state from among these functions.
 
In the literature, there are frequent references to ``chirality.''
Configurations have positive chirality, if when one traverses a triangle
of spins clockwise, the spins rotate clockwise by
$\Delta \phi = 120^{\rm o}$, whereas for negative chirality
$\Delta \phi=-120^{\rm o}$.  In our opinion, the characterization
by irreps is more fundamental.  One sees that the wavefunction of
irrep $\Gamma_1^+$ and one of those of $\Gamma^+_2$ correspond to positive
chirality and two of the wavefunctions for $\Gamma_3^+$ correspond
to negative chirality.  However, the sign of the chirality,
of itself, does not indicate the presence or not of a gap in
the excitation spectrum.  Nor does the sign of the chirality
correlate in an obvious way with the nature of other wavefunctions
which may be admixed.

\subsection{Actual Magnetic Structure}

\begin{figure}
\begin{center}
\includegraphics[height=5cm]{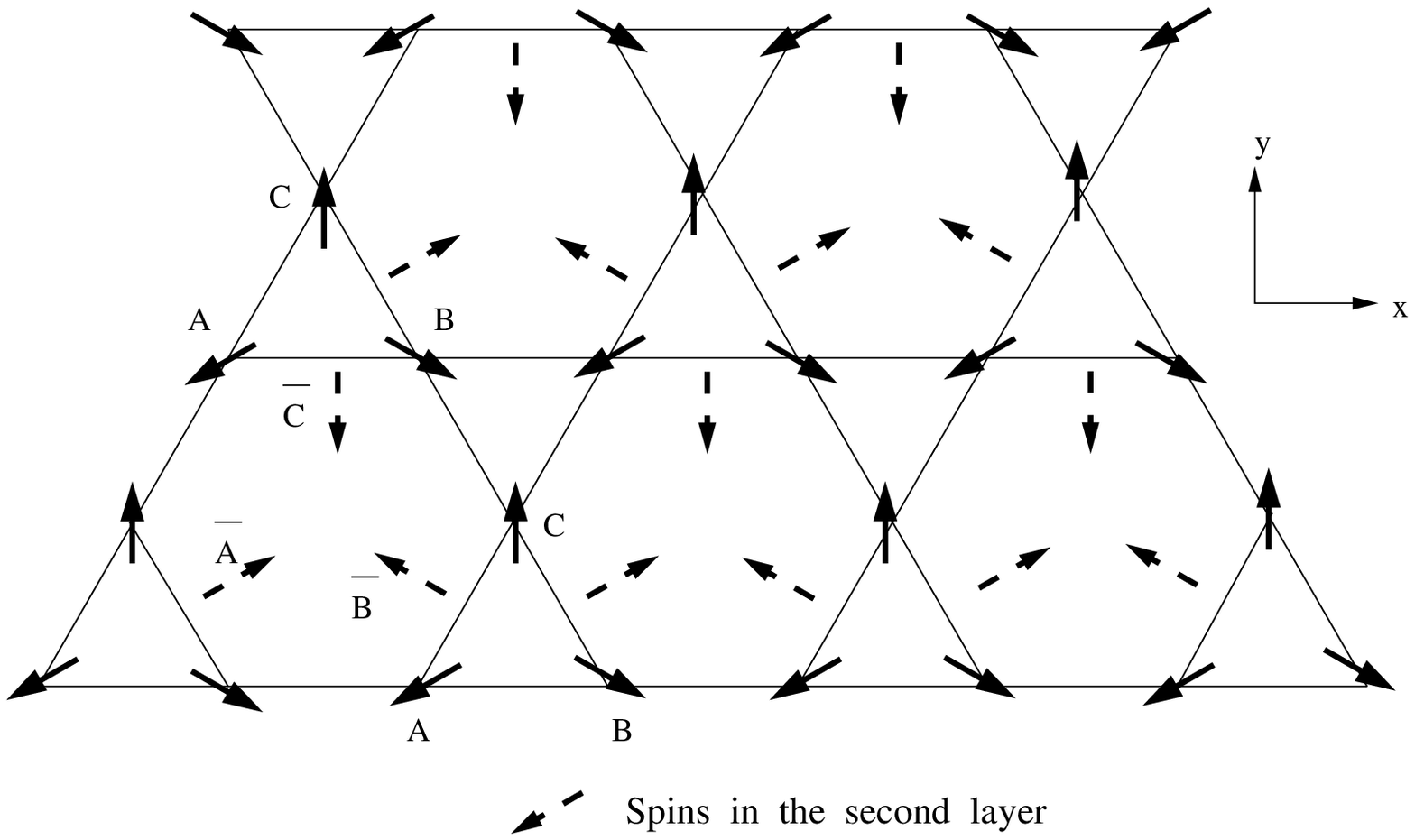} \caption{\label{SPIN}
Spin structure of FeJ as determined in Refs. \protect{\onlinecite{JAPS}},
\protect{\onlinecite{INAMI}}, and \protect{\onlinecite{YLEE1}}.}
\end{center}
\end{figure}

The actual magnetic structure\cite{INAMI}
of FeJ is that shown in Fig. \ref{SPIN}
associated with the wavevector ${\bf q}=3 \pi /c$.  This means that
successive Kagom\'e planes are translated (as they would be in
the paramagnetic phase), but then in successive planes the spin
orientations are reversed.  This causes a doubling of the unit cell
because after three translations, the sites are back to their original
positions, but the spin orientations are reversed.  Thus the magnetic
unit cell contains twice as many layers as in the paramagnetic phase.
Although each layer individually has a small ferromagnetic
moment perpendicular to the Kagom\'e plane, the sign of this
moment alternates in sign from one Kagom\'e net to the next, so that
the sample exhibits no overall net moment.  However, as in
La$_2$CuO$_4$,\cite{LCO} one has a spin-flop transition\cite{YLEE1}
when a magnetic field perpendicular to the Kagom\'e plane rearranges
the planes so that their ferromagnetic moments are parallel.

\begin{figure}
\begin{center}
\includegraphics[height=3.5cm]{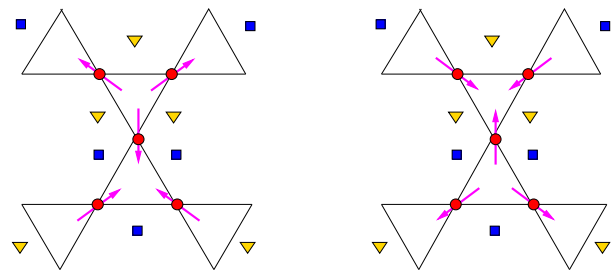} \caption{\label{LRO}
(Color online) Spin structure of FeJ.  The squares (blue) are
sites in the Kagom\'e plane at $z=c/3$, the triangles (yellow)
are sites in the plane at $z=-c/3$ and the circles (red)
are sites in the plane at $z=0$. The two symmetry related choices for
spin orientations in the ordered phase are shown in the two panels.}
\end{center}
\end{figure}

Here we discuss some aspects of broken symmetry.  One could ask whether
the direction of the net moment of a Kagom\'e plane is fixed by
interactions within the plane or is accidentally selected.
First, we should note that if we had a single isolated Kagom\'e
plane, then the two orientations or the net moment perpendicular to the
plane would have identical free energy.  As we will see, the direction
of the net moment is fixed by the sign of the DM interaction.  If
we had a single Kagom\'e plane, such an interaction would not be
allowed.  So, the uniqueness of the direction of the moment perpendicular
to the plane, must depend on the details of the three-dimensional
structure and, because $+z$ and $-z$ are equivalent in the paramagnetic
crystal, on the in-plane magnetic ordering.
To see how this is possible consider Fig. \ref{LRO}.
The actual spin structure is shown in the two panels of Fig. \ref{LRO}.
The selection of one of these states is arbitrary and represents an
example of broken symmetry.  Let us see how the direction of the
net moment perpendicular to the plane is fixed relative to the in-plane
ordering. Suppose we have the
broken symmetry selection of the structure shown in the left panel.
There one sees that the in-plane spin moments point away from sites in the
$z=-c/3$ plane and towards sites in the
$z=c/3$ plane.  This indicates that positive and negative $z$ are
not equivalent {\it due to the presence of long-range magnetic order
in the three-dimensional crystal structure}.
This means that when the spin order is as in the left panel, the
free energy will select the direction of the perpendicular moment.
When the spin order is as in the right panel, the perpendicular moment
will be reversed relative to what it was in the left panel.  As we
will see, when the in-plane order is selected, the  direction of the
out-of plane moment is determined by the sign of the $y$-component
of the DM vector. It is interesting to note that the direction
of the perpendicular moment can not be related to chirality.  The
structures shown in Fig. \ref{LRO} have positive chirality whether seen
by a viewer from in front of the page or behind the page.  So associating
a direction with chirality can not distinguish between directions
into or out of the page. 

\section{Model Hamiltonians}

\subsection{General Hamiltonian}

\begin{figure}
\begin{center}
\includegraphics[height=5cm]{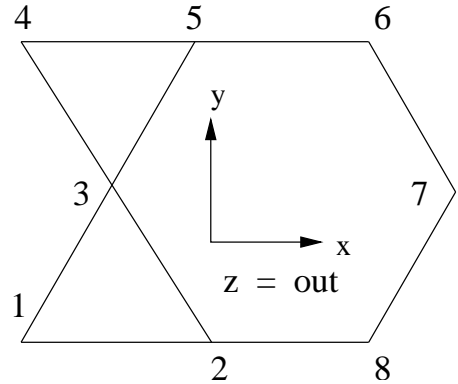} \caption{\label{BONDS}
Numbering of sites used to define nn and nnn interactions.
The axes fixed in the crystal are shown.}
\end{center}
\end{figure}

We neglect interactions between adjacent Kagom\'e planes.  Accordingly,
we now discuss the Hamiltonian of a single Kagom\'e plane.
We first parameterize the nn interaction between spins
\#1 and \#2 in Fig. \ref{BONDS}, which we write as
\begin{eqnarray}
{\cal H}_{12} &=& \sum_{\alpha \beta} M_{1,2}^{\alpha ,\beta}
{\cal S}_{1}^\alpha {\cal S}_{2}^\beta \ ,
\label{DEFEQ} \end{eqnarray}
where the Greek superscripts label Cartesian components.  The
effect of the mirror plane perpendicularly bisecting the 1-2 bond
is to change the sign of the $y$- and $z$-components of spin
and simultaneously interchange spin labels.  Thus taking the
transpose of the matrix and changing the signs of the
$y$- and $z$-components must leave the matrix invariant.  So the
interaction matrix must be of the form
\begin{eqnarray}
{\bf M}_{1,2} &\equiv& {\bf M}^{(1)}= \left[ \begin{array} {c c c}
J_{xx}^{(1)} & D_z^{(1)} & -D_y^{(1)} \\
- D_z^{(1)} & J_{yy}^{(1)} & J_{yz}^{(1)} \\
D_y^{(1)} & J_{yz}^{(1)} & J_{zz}^{(1)} \\ \end{array} \right] \ ,
\label{NN} \end{eqnarray}
where the superscript indicates an nn interaction and
we introduce symmetric anisotropic exchange tensor ${\bf J}$ and antisymmetric
DM interactions,\cite{DM1,DM2} characterized by a DM vector ${\bf D}$.
Note that the definition of Eq. (\ref{DEFEQ}) implies that
\begin{eqnarray}
{\bf M}_{i,j}= \tilde {\bf M}_{j,i} 
\end{eqnarray}
independent of what the local symmetry may be, where tilde indicates
a transposed matrix.

We next characterize the nnn interaction between sites \#5 and \#2
in Fig. \ref{BONDS}.
This interaction has to be invariant under the two-fold rotation about
the $x$-axis passing through the center of the bond.  So the interaction
matrix in Cartesian coordinates must be of the form
\begin{eqnarray}
{\bf M}_{5,2} &\equiv& {\bf M}^{(2)}=  \left[ \begin{array} {c c c}
J_{xx}^{(2)} & D_z^{(2)} & - D_y^{(2)} \\
- D_z^{(2)} & J_{yy}^{(2)} & J_{yz}^{(2)} \\
D_y^{(2)} & J_{yz}^{(2)} & J_{zz}^{(2)} \\ \end{array} \right] \ ,
\label{NNN} \end{eqnarray}
where ${\bf D}^{(2)}$ is the nnn DM interaction and ${\bf J}^{(2)}$
is the symmetric nnn interaction.

We also include a single-ion anisotropy energy $E_i^{(A)}$
for site $i$ of the form
\begin{eqnarray}
E_i^{(A)} &=& (1/2) \sum_{\alpha \beta} [{\bf C}']_i^{\alpha \beta}
[S_i^\alpha S_i^\beta + S_i^\beta S_i^\alpha ] \ .
\end{eqnarray}
The mirror $x$-plane through site \#3 implies that the
single-ion anisotropy matrix for this site is
\begin{eqnarray}
{\bf C}_3^\prime &=& \left[ \begin{array} {c c c}
C_{xx}^\prime & 0 & 0 \\
0 & C_{yy}^\prime & C_{yz}^\prime \\ 
0 & C_{yz}^\prime & C_{zz}^\prime \\ \end{array} \right] \ .
\end{eqnarray}

\subsection{Transformation to Local Uncanted Axes}

\begin{figure}
\begin{center}
\includegraphics[height=5cm]{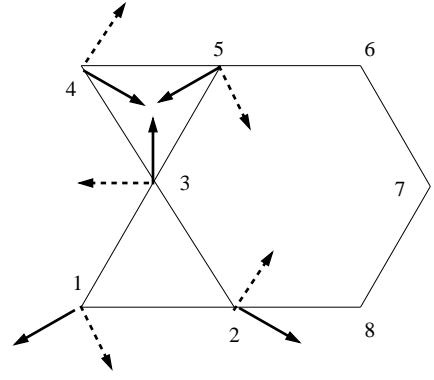} \caption{\label{LOCAL}
Local axes of various sites.  The local $z$-axis (indicated
by the full line with arrow) is along the projection of the
direction of the local magnetization onto the Kagom\'e plane.
The local $y$-axis is perpendicularly out of the Kagom\'e
plane. The local $x$-axis is in the Kagom\'e plane and is
indicated by the dashed line with arrow.}
\end{center}
\end{figure}

A direct, but inefficient, procedure would be to use the symmetry
of the crystal to express all the other nn and nnn interactions in
terms of the matrices of Eqs. (\ref{NN}) and (\ref{NNN}), respectively.
Instead, we proceed as follows.
We express the nn and nnn interactions in terms of local
axes which facilitate a spin-wave expansion.  These local axes
are defined so that the local positive $z$ axis coincides with the
projection of the local spin moment onto the Kagom\'e plane.
These axes are illustrated in Fig. \ref{LOCAL}.  The  rotation
matrices to transform to these local axes are
\begin{eqnarray}
{\cal R}(\tau_1) &=& \left[ \begin{array} {c c c}
{1 \over 2} & 0 & - {\sqrt 3 \over 2} \\ - {\sqrt 3 \over 2} & 0 & - {1 \over 2} \\
0 & 1 & 0 \\ \end{array} \right] \ , \nonumber \\
{\cal R}(\tau_2) &=& \left[ \begin{array} {c c c}
{1 \over 2} & 0 & {\sqrt 3 \over 2} \\  {\sqrt 3 \over 2} & 0 & - {1\over 2} \\
0 & 1 & 0 \\ \end{array} \right] \ , \nonumber \\
{\cal R}(\tau_3) &=& \left[ \begin{array} {c c c}
-1 & 0 & 0 \\  0 & 0 & 1 \\ 0 & 1 & 0 \\ \end{array} \right] \ .
\end{eqnarray}
The transformed interaction matrices, denoted ${\cal I}$, are such that
\begin{eqnarray}
{\cal H}_{12} &=& \sum_{\alpha \beta} {\cal I}_{1,2}^{(\alpha, \beta)}
S_1^\alpha S_2^ \beta \ ,
\end{eqnarray}
and
\begin{eqnarray}
{\cal H}_{52} &=& \sum_{\alpha \beta} {\cal I}_{5,2}^{(\alpha, \beta)}
S_5^\alpha S_2^ \beta \ ,
\end{eqnarray}
where now the spin components are taken in the local axes of Fig.
\ref{LOCAL} so that
\begin{eqnarray}
{\cal I}_{1,2} &=& \tilde {\cal R}(\tau_1) {\bf M}_{1,2} {\cal R}(\tau_2)
\equiv {\cal I}^{(1)} \nonumber \\
{\cal I}_{5,2} &=& \tilde {\cal R}(\tau_1) {\bf M}_{5,2} {\cal R}(\tau_2)
\equiv {\cal I}^{(2)} \ . \end{eqnarray}
Here
\begin{eqnarray} 
{\cal I}^{(n)} &=& \left[ \begin{array} {c c c}
-E_x^{(n)} & d_z^{(n)} & -d_y^{(n)} \\ -d_z^{(n)} &
- E_y^{(n)} & E_{yz}^{(n)} \\
d_y^{(n)} & E_{yz}^{(n)} & -E_z^{(n)} \\ \end{array}
\right] \ ,
\label{EFFSW} \end{eqnarray}
where
\begin{eqnarray}
E_x^{(n)} &=& {3 \over 4} J_{yy}^{(n)} - {1 \over 4} J_{xx}^{(n)}
- {\sqrt 3 \over 2} D_z^{(n)} \nonumber \\ &\equiv &
{1 \over 2} J_n - {\sqrt 3 \over 2} D_z^{(n)} - {7 \over 24} \Delta_n 
- {1 \over 8} \eta_n
\nonumber \\ E_y^{(n)} &=& - J_{zz}^{(n)}  \nonumber \\
&\equiv& -J_n - {1 \over 6} \Delta_n + {1 \over 2} \eta_n \nonumber \\
E_z^{(n)} &=& {3 \over 4} J_{xx}^{(n)} - {1 \over 4} J_{yy}^{(n)}
- {\sqrt 3 \over 2} D_z^{(n)} \nonumber \\ &\equiv&
{1 \over 2} J_n - {\sqrt 3 \over 2} D_z^{(n)} + {5 \over 24} \Delta_n +
{3 \over 8} \eta_n \nonumber \\
d_z^{(n)} &=& - {1 \over 2} D_y^{(n)} - {\sqrt 3 \over 2} J_{yz}^{(n)}
\nonumber \\
E_{yz}^{(n)} &=& {\sqrt 3 \over 2} D_y^{(n)} - {1 \over 2} J_{yz}^{(n)}
\nonumber \\
d_y^{(n)} &=& - {\sqrt 3 \over 4} (J_{xx}^{(n)} + J_{yy}^{(n)})
- {1 \over 2} D_z^{(n)} \nonumber \\ &\equiv&
- {\sqrt 3 \over 2} J_n - {1 \over 2} D_z^{(n)} + {\sqrt 3 \over 4} \left(
{\Delta_n \over 6} - {\eta_n \over 2} \right) \  ,
\label{PARAM} \end{eqnarray}
where
\begin{eqnarray}
J_n &=& (J_{xx}^{(n)} + J_{yy}^{(n)} + J_{zz}^{(n)})/ 3 \nonumber \\
\Delta_n &=& J_{xx}^{(n)} + J_{zz}^{(n)} - 2 J_{yy}^{(n)} \ ,  \ \ \
\eta_n = J_{xx}^{(n)} - J_{zz}^{(n)} \ ,
\end{eqnarray}
so that $J_n$ is the isotropic part of the $n$th neighbor interaction
and we will only keep the anisotropic nn interactions, so that henceforth
$\Delta_J \equiv \Delta_1$ and $\eta_J \equiv \eta_1$.
We define $E_\alpha^{(n)}$ with a sign so that the interactions
appear ferromagnetic in the local frame. The matrix elements 
$E_{yz}^{(n)}$ lead to a canted
spin structure exhibiting weak ferromagnetism, as we will see 
in the next subsection.
The single-ion anisotropy in the local frame can be written as
\begin{eqnarray}
{\bf C} &=& \left[ \begin{array} {c c c}
C_{xx} & 0 & 0 \\
0 & C_{yy} & C_{yz} \\ 
0 & C_{yz} & C_{zz} \\ \end{array} \right] \ .
\label{CPARAM} \end{eqnarray}

A big advantage of expressing the interactions in terms of
local axes is that all single-ion anisotropies and all
nn interaction matrices are identical
(apart from transposing if site indices are interchanged).
Using the three-fold axis and inversion symmetry we find that
\begin{eqnarray}
{\cal I}_{1,2} &=& {\cal I}_{2,3} = {\cal I}_{3,1}
= \tilde {\cal I}_{2,1} = \tilde {\cal I}_{3,2} = \tilde {\cal I}_{1,3} 
\nonumber \\ &=& {\cal I}_{4,3} =  {\cal I}_{5,4} = {\cal I}_{3,5} 
= \tilde {\cal I}_{3,4} = \tilde {\cal I}_{4,5} = \tilde {\cal I}_{5,3} \ .
\label{SYMMETRY} \end{eqnarray}
where the sites are numbered as in Fig. \ref{LOCAL}.

Just as for nearest neighbors, all the next nearest neighbor
interaction matrices are the same, providing we take the indices in the
correct order. 

\subsection{Canting}

Up to now we have taken the local $z$-axes to lie in the
Kagom\'e plane.  However, the matrix elements $E_{yz}^{(n)}$
for $n=1$ or $n=2$ gives rise to a field which induces a
uniform $y$ component of spin.
For the spin-wave calculation it is convenient to define
``canted local axes'' such that the canted $z$-axes lie along
the direction of the canted spins. This direction is found
by minimizing the energy.  We write the
ground state energy per site, $E_G$,
for $S_z=S \cos \theta$ and $S_y=S \sin \theta$ as
\begin{eqnarray}
E_G/S^2 &=& [ C_{zz} - 2E_z^{(1)}-2E_z^{(2)}] \cos^2 \theta
\nonumber \\ && \ + [ C_{yy} - 2 E_y^{(1)}-2E_y^{(2)} ] \sin^2 \theta
\nonumber \\ && \ + 2 [ C_{yz} + 2 E_{yz}^{(1)}+2E_{yz}^{(2)}]
\sin \theta \cos \theta \nonumber \\ & \equiv &
- A_{cc} \cos^2 \theta - A_{ss} \sin^2 \theta 
+ 2 A_{cs} \sin \theta \cos \theta \nonumber \\
&=& - {1 \over 2} [A_{cc}+A_{ss}]
- H_z \cos (2 \theta) \nonumber \\ && \ - H_y \sin (2 \theta ) \ ,
\end{eqnarray}
where 
\begin{eqnarray}
H_z = [A_{cc}-A_{ss}]/2 \ , \ \ \ \ H_y =  - A_{cs} \ . 
\end{eqnarray}
The energy is minimized by setting
\begin{eqnarray}
\cos (2 \theta) &=& H_z/H \ , \ \ \ \ \sin (2 \theta) = H_y/H \ ,
\end{eqnarray}
where $H= \sqrt{H_z^2 + H_y^2}$.  If we neglect the effect of anisotropic
nnn interactions on the canting angle, we have
\begin{eqnarray}
H_y &=& -[C_{yz} + 2E_{yz}^{(1)} ]
= -C_{yz} - \sqrt 3 D_y + J_{yz} \nonumber \\
H_z &=& E_z^{(1)} - E_y^{(1)} - {1 \over 2} C_{zz}
+ {1 \over 2} C_{yy} \nonumber \\ && \ 
+ E_z^{(2)} - E_y^{(2)} \nonumber \\ 
&=& {3 \over 2} J_1 - {\sqrt 3 \over 2} D_z + {C_{y-z} \over 2}
+ {3 \over 2} J_2 + {3 \Delta_J - \eta_J \over 8}  \ ,
\label{HYZEQ} \end{eqnarray}
where $C_{\alpha-\beta}\equiv C_{\alpha \alpha} - C_{\beta \beta}$.
Here and below the superscript ``1'' for nn's is implied in
the anisotropic interactions.

\begin{center}
\begin{figure}
\includegraphics[scale=0.8]{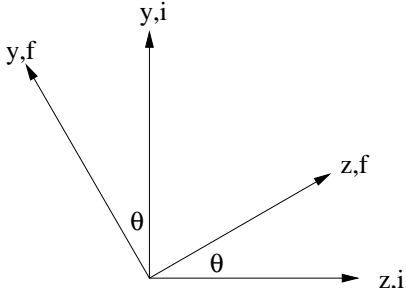}
\caption{\label{AXES}
The `initial' and `final' local $y$ and $z$ axes.  Initially
the $z$ axis is in the Kagom\'e plane.  The sign of $\theta$ is
the same that of $H_y$, which is the negative of $E_{yz}^{(1)}$, i .e.
the negative of $D_y$.}
\end{figure}
\end{center}

\noindent
This means that we want to transform the initial local coordinates
with moments in the Kagom\'e plane into the canted local coordinates via
\begin{eqnarray}
S_z^{\rm loc,i} &=& c S_z^{\rm loc,f} - s S_y^{\rm loc,f}
\nonumber \\
S_y^{\rm loc,i} &=& s S_z^{\rm loc,f} + c S_y^{\rm loc,f}
\ ,
\end{eqnarray}
as illustrated in Fig. \ref{AXES}, where $c \equiv \cos \theta$ and
$s\equiv  \sin \theta$.  Then, if we include this transformation we finally
arrive at the local anisotropy matrix as
\begin{eqnarray}
\overline {\bf C} &=& \left[ \begin{array} {c c c} C_{xx} & 0 & 0 \\
0 & \overline C_{yy} & \overline C_{yz}  \\ 0 & \overline C_{yz} &
\overline C_{zz} \\ \end{array} \right] \ ,
\label{CDEF} \end{eqnarray}
and the transformed local interaction matrices are
\begin{eqnarray}
\overline {\cal I}_{1,2} &=& \left[ \begin{array} {c c c}
-E_x^{(1)} & \overline d_z^{(1)} & - \overline d_y^{(1)} \\
- \overline d_z^{(1)} & - \overline E_y^{(1)} &
\overline E_{yz}^{(1)} \\ \overline d_y^{(1)}
& \overline E_{yz}^{(1)} & - \overline E_z^{(1)} \\
\end{array} \right] \ ,
\end{eqnarray}
and similarly for $\overline {\cal I}_{5,2}$, where
\begin{eqnarray}
\overline C_{zz} &=& c^2 C_{zz} + s^2 C_{yy} + 2cs C_{yz} \ , \nonumber \\
\overline E_z^{(n)} &=& c^2 E_z^{(n)} + s^2 E_y^{(n)} - 2cs E_{yz}^{(n)} \ ,
\nonumber \\
\overline C_{yy} &=& s^2 C_{zz} + c^2 C_{yy} - 2cs C_{yz} \ , \nonumber \\
\overline E_y^{(n)} &=& s^2 E_z^{(n)} + c^2 E_y^{(n)} + 2cs E_{yz}^{(n)} \ ,
\nonumber \\
\overline C_{yz} &=& C_{yz} \cos (2 \theta) + {1 \over 2}
[ C_{yy} - C_{zz} ] \sin ( 2 \theta) \ , \nonumber \\
\overline E_{yz}^{(n)} &=& E_{yz}^{(n)} \cos (2 \theta) +
{1 \over 2} [E_z^{(n)} - E_y^{(n)}] \sin(2 \theta) \nonumber \\
\overline d_z^{(n)} &=& c d_z^{(n)} + s d_y^{(n)} \nonumber \\
\overline d_y^{(n)} &=& c d_y^{(n)} - s d_z^{(n)} \ .
\label{TILDE} \end{eqnarray}
An equivalent condition for equilibrium is that
\begin{eqnarray}
\overline C_{yz} + 2 \sum_n \overline E_{yz}^{(n)} = 0 \ .
\end{eqnarray}
When the $x$-$z$ elements are summed over all neighbors they give
a vanishing effective field.  Indeed, group theory
guarantees that this is the case for general interactions.

\section{Calculation of the Spin Wave Spectrum}

In Appendix \ref{BOSON} we write the Hamiltonian in terms of boson
operators and in Appendix \ref{MODES} we obtain the wavevector
dependent spin-wave dynamical matrix (SWDM) ${\bf M}({\bf q})$
whose eigenvalues are the spin-wave energies at wavevector ${\bf q}$
and which is written as
\begin{eqnarray}
{\bf M}({\bf q}) &=& \left[ \begin{array} {c | c} \hline
{\bf A}({\bf q}) & -{\bf B}({\bf q}) \\ \hline
{\bf B}({\bf q})^* & -{\bf A}({\bf q})^* \\ \hline
\end{array} \right] \ .
\end{eqnarray}

\subsection{General Form of the Spin-Wave Dynamical Matrix}

The matrices ${\bf A}({\bf q})$ and ${\bf B}({\bf q})$ are given
in Eqs. (\ref{ADEF}) and (\ref{BDEF}), respectively.  We write them as
\begin{eqnarray}
{\bf A}({\bf q}) &=& \overline \Delta {\cal E} + \sum_n {\bf A}^{(n)}
({\bf q})  
\label{AMATEQ} \end{eqnarray}
and
\begin{eqnarray}
{\bf B}({\bf q}) &=& (C_{xx}- \overline C_{yy}) {\cal E} 
+ \sum_n {\bf B}^{(n)} ({\bf q}) \ ,  
\end{eqnarray}
where $\overline \Delta= C_{xx}+ \overline C_{yy}- 2\overline C_{zz}$,
${\cal E}$ is the unit matrix, and for $n=1$ or 2 ${\bf A}^{(n)}({\bf q})$ is
\begin{small}
\begin{eqnarray}
\left[ \begin{array} {c | c | c} \hline
4\overline E_z^{(n)} & [-E_x^{(n)} - \overline E_y^{(n)}
& [-E_x^{(n)} - \overline E_y^{(n)} \\
& -2 i \overline d_z^{(n)}] \gamma_{1,2}^{(n)}({\bf q}) &
+ 2 i \overline d_z^{(n)}] \gamma_{1,3}^{(n)}({\bf q}) \\
\hline
[-E_x^{(n)} - \overline E_y^{(n)}
& 4\overline E_z^{(n)}  & [-E_x^{(n)} - \overline E_y^{(n)} \\
+ 2 i \overline d_z^{(n)}] \gamma_{1,2}^{(n)}({\bf q}) & 
& -2 i \overline d_z^{(n)}] \gamma_{2,3}^{(n)}({\bf q}) \\
\hline [-E_x^{(n)} - \overline E_y^{(n)}
& [-E_x^{(n)} - \overline E_y^{(n)} & 4\overline E_z^{(n)} \\
-2 i \overline d_z^{(n)}] \gamma_{1,3}^{(n)}({\bf q})
&+2 i \overline d_z^{(n)}] \gamma_{2,3}^{(n)}({\bf q}) & \\
\hline \end{array} \right] \ ,
\label{MATAEQ} \end{eqnarray}
\end{small}
and ${\bf B}^{(n)}({\bf q})$ is
\begin{small}
\begin{eqnarray}
\left[ \begin{array} {c | c | c} \hline 0
& [ - E_x^{(n)} + \overline E_y^{(n)}] &
[ - E_x^{(n)} + \overline E_y^{(n)}] \\
& \times \gamma_{1,2}^{(n)}({\bf q}) & \times \gamma_{1,3}^{(n)}({\bf q})
\\ \hline
[ - E_x^{(n)} + \overline E_y^{(n)}] & 0 &
[ - E_x^{(n)} + \overline E_y^{(n)}] \\
\times \gamma_{1,2}^{(n)}({\bf q}) & & \times \gamma_{2,3}^{(n)}({\bf q})
\\ \hline [ - E_x^{(n)} + \overline E_y^{(n)}] &
[ - E_x^{(n)} + \overline E_y^{(n)}] & 0 \\
\times \gamma_{1,3}^{(n)}({\bf q}) & \times \gamma_{2,3}^{(n)}({\bf q}) &
\\ \hline \end{array} \right] \ ,
\label{MATBEQ} \end{eqnarray}
\end{small}
where $\gamma_{\tau,\tau'}^{(n)}({\bf q})$ is the normalized form factor
for the $n$th shell of neighbors.  Here we we will only treat $n=1,2$.
For in-plane interactions we have
\begin{eqnarray}
\gamma_{12}^{(1)}({\bf q}) &=& \cos (a q_x/2) \ , \nonumber \\
\gamma_{13}^{(1)}({\bf q}) &=& \cos [a (q_x + \sqrt 3 q_y)/4] \ ,
\nonumber \\
\gamma_{23}^{(1)}({\bf q}) &=& \cos [a (q_x - \sqrt 3 q_y)/4] \ ,
\nonumber \\
\gamma_{12}^{(2)}({\bf q}) &=& \cos (a q_y \sqrt 3 /2) \ , \nonumber \\
\gamma_{13}^{(2)}({\bf q}) &=& \cos [a (3 q_x - \sqrt 3 q_y)/4] \ ,
\nonumber \\
\gamma_{23}^{(2)}({\bf q}) &=& \cos [a (3 q_x + \sqrt 3 q_y)/4] \ .
\end{eqnarray}
In Eqs. (\ref{MATAEQ}) and (\ref{MATBEQ}) one
one uses Eq. (\ref{TILDE}) to relate the overlined coefficients
to the bare coefficients, which are defined in Eqs. (\ref{CDEF})
and (\ref{PARAM}). 

To simplify the expressions we will now specialize to the case when
the nnn interactions are isotropic and all
interactions further than nnn are ignored.
Then we write the matrix ${\bf A}({\bf q})$ in the form
\begin{small}
\begin{eqnarray}
\left[ \begin{array} {c | c | c} \hline
A_0 + 2J_2 & (A_1- i \alpha_1) \gamma_{12}^{(1)} +
& (A_1+i\alpha_1) \gamma_{13}^{(1)} + \\
& \left({1 \over 2}J_2-i\alpha_2\right) \gamma_{12}^{(2)}&
\left( {1 \over 2}J_2 -i\alpha_2 \right) \gamma_{13}^{(2)} \\ 
\hline
(A_1 + i \alpha_1 ) \gamma_{12}^{(1)} +
& A_0+ 2J_2 & (A_1 - i \alpha_1) \gamma_{23}^{(1)} + \\ 
\left( {1 \over 2} J_2 + i \alpha_2 \right) \gamma_{12}^{(2)} & &
\left( {1 \over 2} J_2 - i \alpha_2 \right) \gamma_{23}^{(2)} \\
\hline (A_1 - i \alpha_1) \gamma_{13}^{(1)} + &
(A_1+ i \alpha_1) \gamma_{23}^{(1)} + & A_0+2J_2 \\
\left( {1 \over 2} J_2 - i \alpha_2 \right) \gamma_{13}^{(2)} &
\left( {1 \over 2} J_2 + i \alpha_2 \right) \gamma_{23}^{(2)} & \\
\hline \end{array} \right] \ ,
\label{NMATAEQ} \end{eqnarray}
\end{small}
where $\gamma_{nm}^{(k)}$ denotes $\gamma_{nm}^{(k)}({\bf q})$. 
The matrix ${\bf B}({\bf q})$ is of the form
\begin{small}
\begin{eqnarray}
\left[ \begin{array} {c | c | c} \hline
B_0 & B_1 \gamma_{12}^{(1)}({\bf q}) &
B_1 \gamma_{13}^{(1)}({\bf q}) \\
& - {3 \over 2} J_2 \gamma_{12}^{(2)}({\bf q}) &
- {3 \over 2} J_2 \gamma_{13}^{(2)}({\bf q}) \\ \hline
B_1 \gamma_{12}^{(1)}({\bf q}) & B_0 &  B_1 \gamma_{23}^{(1)}({\bf q}) \\
- {3 \over 2} J_2 \gamma_{12}^{(2)}({\bf q}) & &
- {3 \over 2} J_2 \gamma_{23}^{(2)}({\bf q}) \\ \hline
B_1 \gamma_{13}^{(1)}({\bf q}) & B_1 \gamma_{23}^{(1)}({\bf q}) & B_0 \\
- {3 \over 2} J_2 \gamma_{13}^{(2)}({\bf q}) &
- {3 \over 2} J_2 \gamma_{23}^{(2)}({\bf q}) &
\\ \hline \end{array} \right] \ .
\label{NMATBEQ} \end{eqnarray}
\end{small}
In Appendix \ref{EVALUATE} we evaluate of the constants in the
matrices for the general Hamiltonian introduced above. In the
interest of simplicity we now limit consideration to
the following generic Hamiltonian:

\begin{eqnarray}
{\cal H} &=& \sum_{i<j \in nn} \biggl[ J_1 S_i \cdot S_j
+ D_{ij}\cdot S_i \times S_j \biggr]
\nonumber \\ && \ +\sum_{k<l \in nnn} J_2 S_k \cdot S_l
\nonumber \\
& +& D \sum_{i} (S_i^{y'})^2 - E \sum_{i} [ (S_i^{z'})^2 -(S_i^{x'})^2 ]
\label{model_H}
\end{eqnarray} 
where $\in nn$ ($\in nnn$) indicates that the sum is over nn's (nnn's),
$D_{ij} = (0,D_y(i,j),D_z(i,j))$ is the Dzyaloskinskii-Moriya vector
for bond $i-j$ as shown in Fig. \ref{BONDS}  and the single-ion anisotropy
terms are those used by Nishiyama {\it et al.}\cite{JAPS} in their treatment
of the spin-wave spectrum in jarosites.  Here the prime-spin components
refer to the local axis associated with the rotated oxygen octahedra shown
in Fig. 1 (see Ref. \onlinecite{JAPS} for details).  The $D_y(i,j)$
and $D_z(i,j)$ are all expressible in terms of the parameters $D_y$
and $D_z$ as discussed in Sec. III.  The single-ion anisotropy constants
$D$ and $E$ used in Ref. \onlinecite{JAPS} are related to our
generic single-ion matrix $C$ defined in the uncanted-local frame given
in Eq. (\ref{CPARAM}) as
\begin{eqnarray}
C_{xx} &=& E  \nonumber \\
C_{xy} &= &C_{xz} = C_{yx} = C_{zx} = 0 \nonumber \\
C_{yy} &=&  D \cos(\theta_0)^2 - E \sin(\theta_0)^2 \nonumber  \\
C_{yz} &=& C_{zy} = (D+E)\sin(\theta_0)\cos(\theta_0) \nonumber \\ 
C_{zz} &=& D \sin(\theta_0)^2 - E \cos(\theta_0)^2  \ ,
\label{DE_relation} \end{eqnarray}
where $\theta_0$ is the rotation angle of the octahedra around Fe-ion
(see Fig. \ref{ALUNITE}) and is $\theta_0= 20^{\rm o}$.\cite{JAPS}

In what follows we will often have recourse to two simple models.
In the first of these, which we call the ``DM'' model, we neglect the
single ion anisotropy and the exchange anisotropy, so that the only
nonzero parameters are $J_1$, $J_2$, $D_y$, and $D_z$.  In the
second model, which we call the ``CF'' model, all the anisotropy
is incorporated by the single-ion crystal field, so that the only
nonzero parameters are $J_1$, $J_2$, and $C_{\alpha \beta}$.
In either case, we assume $J_1$ to be the dominant interaction.
Accordingly we give here the expressions for the matrix elements of
the SWDM which pertain to these two cases:
\begin{eqnarray}
A_0 &=& 2J_1 + \Delta - 2 \sqrt 3 D_z
+ {2  \over 3J_1} [3 D_y^2 + 2 C_{yz}^2] \ , \nonumber \\
A_1 &=& {1 \over 2} J_1 + {\sqrt 3 \over 2} D_z
+ {1 \over 6J_1} \biggl[ 3D_y^2-C_{yz}^2 \biggr] \ , \nonumber \\
B_0 &=& C_{xx}-C_{yy}  - {2C_{yz}^2 \over 3J_1} \ , \nonumber \\ 
B_1 &=& - {3 \over 2} J_1 + {\sqrt 3 \over 2} D_z 
+ {C_{yz}^2 - 3 D_y^2 \over 6J_1} \ , \nonumber \\
\alpha_1 &=& {1 \over \sqrt 3} C_{yz}
+ {2 \sqrt 3  D_y D_z \over 3J_1} - {D_y J_2 \over J_1} \ , \nonumber \\
\alpha_2 &=& {D_y J_2 \over J_1} \ .
\label{EQ47}
\end{eqnarray}
where $\Delta = C_{xx}+C_{yy}-2C_{zz}$.

\subsection{Analytic Results for Zero Wavevector}

At zero wavevector the SWDM,
${\bf M}$, whose eigenvalues give $(\omega/S)\equiv \tilde \omega$, is
\begin{table*}
\caption{\label{OMEGA} Spin wave energies $\tilde \omega \equiv \omega/S$
for zero wavevector for the DM model (only $J_1$, $J_2$, $D_y$, and $D_z$
nonzero) and for the CF model (only $J_1$, $J_2$ and $C_{\alpha \beta}$
nonzero). The results neglecting canting are discussed in subsection IVE.
Here $J\equiv J_1+J_2$. In this and succeeding tables $C_{\alpha-\beta}$
denotes $C_{\alpha\alpha}-C_{\beta \beta}$.}

\vspace{0.2 in}
\begin{tabular} { | c | c | c ||}
\hline \hline
\multicolumn {3} {|c|}  {With Canting} \\ \hline
& DM Model & CF Model \\ \hline
$\tilde \omega_0$ &
$\sqrt {12}|D_y| + {\cal O}(D^2/J)$ &
$\left[ 12JC_{x-z}+4C_{x-z}C_{y-z}
+ 4C_{yz}^2 \right]^{1/2} + {\cal O}(C^{5/2}/J^{3/2})$ \\ 
${\tilde \omega_+ + \tilde \omega_-\over 2}$ &
$\sqrt{-6 \sqrt 3 JD_z + 3D_y^2 +18D_z^2}+ {\cal O}(D^{5/2}/J^{3/2})$
& $\left[ 6JC_{y-z} + 4C_{x-z}C_{y-z}
+ 7C_{yz}^2 \right]^{1/2} + {\cal O}(C^{5/2}/J^{3/2})$ \\ 
$\tilde \omega_+-\tilde \omega_-$ & $4D_yD_z/J + {\cal O}(D^3/J^2)$ &
$2C_{yz} + {\cal O} (C^2/J)$ \\
\hline \hline
\multicolumn {3} {|c|} {Neglecting Canting} \\ \hline
& DM Model & CF Model \\ \hline
$\tilde \omega_0$ & $0$
& $\left[ 12JC_{x-z} + 4C_{x-z}C_{y-z}\right]^{1/2}$ \\
${\tilde \omega_+ +\tilde \omega_-\over 2}$ & 
$\sqrt{-6 \sqrt 3 JD_z +18D_z^2}$ &
$\left[ 6JC_{y-z} + 4C_{x-z}C_{y-z} \right]^{1/2}$ \\ 
$\tilde \omega_+ -\tilde \omega_-$ & $2 \sqrt 3 D_y$ & $0$  \\
\hline \hline
\end{tabular}
\end{table*}
 
\begin{tiny}
\begin{eqnarray}
\left[ \begin{array} {c c c | c c c}
A_0 & A_1 - i \alpha & A_1 + i \alpha & -B_0 & - B_1 & - B_1 \\
A_1 + i \alpha & A_0 & A_1-i\alpha & -B_1 & - B_0 & -B_1 \\
A_1 - i \alpha & A_1 + i \alpha & A_0 & -B_1 & -B_1 & -B_0 \\ \hline
B_0 & B_1 & B_1 & - A_0 & -A_1-i \alpha & - A_1 + i \alpha \\
B_1 & B_0 & B_1 & -A_1+i \alpha & -A_0 & -A_1 - i \alpha \\
B_1 & B_1 & B_0 & -A_1-i \alpha & -A_1 +i \alpha & -A_0 
\end{array} \right] \ , \nonumber \\ && 
\label{GMEQ}
\end{eqnarray}
\end{tiny}
where $\alpha=\alpha_1+\alpha_2$ and we include the effects of
$J_2$ by replacing $J_1$ in Eq. (\ref{EQ47}) everywhere by $J_1+J_2$.
Take $|1\rangle$ to have components $(1,1,1,0,0,0)/\sqrt 3$ and
$|2 \rangle$ to have components $(0,0,0,1,1,1)/\sqrt 3$.  Then
\begin{eqnarray}
{\bf M} |1 \rangle &=& (A_0 + 2A_1) |1 \rangle + (B_0+2B_1) |2 \rangle \nonumber \\
{\bf M} |2 \rangle &=& (-B_0 - 2B_1) |1 \rangle + (-A_0-2A_1) |2 \rangle \ .
\end{eqnarray}
From this we find the spin-wave energy $\tilde \omega_0$ to be
\begin{eqnarray}
(\tilde \omega_0)^2 = (A_0 + 2 A_1)^2 - (B_0+2B_1)^2 \ .
\end{eqnarray}

Now take $|3\rangle$ to have components $(2,-1,-1,0,0,0)/\sqrt 6$,
$|4\rangle$ to have components $(0,0,0,2,-1,-1)/\sqrt 6$,
$|5\rangle$ to have components $(0,1,-1,0,0,0)/\sqrt 2$,
and $|6\rangle$ to have components $(0,0,0,0,1,-1)/\sqrt 2$.
In this subspace
\begin{eqnarray}
{\bf M} &=& \left[ \begin{array} {c c c c}
a & b & i \sqrt 3 \alpha & 0 \\
-b & -a & 0 & i \sqrt 3 \alpha \\
-i \sqrt 3 \alpha & 0 & a & b \\
0 & -i \sqrt 3 \alpha & -b & - a \\ \end{array} \right] \ ,
\end{eqnarray}
where $a=A_0 - A_1$, $b= B_0 - B_1$, and $\alpha=\alpha_1+\alpha_2$.
Thereby we find that the other two spin-wave energies, 
$\tilde \omega_\pm$, are given by
\begin{eqnarray}
\tilde \omega_\pm &=& \sqrt{a^2 - b^2} \pm \sqrt 3 \alpha \ .
\end{eqnarray}
Thus for zero wavevector we find the frequencies $\tilde \omega$ 
for the DM and CF models to be those given in Table \ref{OMEGA}
(under the heading "With Canting").

When $\alpha$ is nonzero, we split the degeneracy between the
two heretofore degenerate frequencies.  This splitting is linear in
$\alpha$.  In contrast, when $\alpha=0$, the splitting between
this two-fold degenerate mode and the other mode will be found to be
of order $\sqrt {J_1 \delta}$, where $\delta$ involves anisotropic
exchange or DM interactions.

Note that for ${\bf C}=0$, we have stability for $D_z < D_{zc}$, where
\begin{eqnarray}
{D_{zc} \over J} &=& {\sqrt 3 \over 6} \left( {D_y \over J} \right)^2 
+ {\cal O}[(D_y/J)^4]\ ,
\label{STABLE} \end{eqnarray}
which is similar to Elhajal et al\cite{ELHAJAL} who give
$D_{zc}/J \approx 0.22(D_z/J)^2$. For ${\bf D}=0$, we have stability
(in the large $J$ limit) if both $C_{xx}-C_{zz}$ and $C_{yy}-C_{zz}$
are positive, so that the $z$-axis is really the easiest axis.

\subsection{Results for Other Wavevectors}

\subsubsection{The X Point}
The X point is at
\begin{eqnarray}
q_x &=& [4 \pi / (3a)] \ , \ \ \ \ q_y = q_z=0 \
\end{eqnarray}
and the form factors are
\begin{eqnarray}
\gamma_{1,2}^{(1)} &=& - \oh \ , \ \ \ \
\gamma_{2,3}^{(1)} = \oh \ , \ \ \ \
\gamma_{1,3}^{(1)} = \oh \ ,
\end{eqnarray}
and
\begin{eqnarray}
\gamma_{1,2}^{(2)} &=& 1 \ , \ \ \ \
\gamma_{2,3}^{(2)} = -1 \ , \ \ \ \
\gamma_{1,3}^{(2)} = -1 \ .
\end{eqnarray}

\begin{table*}
\caption{\label{OMEGAX}Spin wave energies $\tilde \omega \equiv \omega/S$
at the X point for the DM and CF models.  Here $J_2$ is considered to be of
the same order as $D$ or $C$. The results neglecting canting are discussed
in Sec. IVE.}

\vspace{0.2 in}
\begin{tabular} { | c | c | c ||}
\hline \hline
\multicolumn {3} {|c|}  {With Canting} \\ \hline
& DM Model & CF Model \\ \hline
$\tilde \omega_0$ &
$\left[6J_1(3J_2-\sqrt 3 D_z) +3 D_y^2 -18\sqrt 3 J_2D_z\right.$ &
$\left[ 6J_1(3J_2+C_{y-z}) + 4C_{x-z}C_{y-z} +7C_{yz}^2 \right.$ \\
& $\left. + 18D_z^2\right]^{1/2} + {\cal O} (D^{5/2}/J^{3/2})$ &
$\left. + 12J_2C_{x-z} \right]^{1/2} + {\cal O}(C^{5/2}/J^{3/2})$ \\ 
${\tilde \omega_+ + \tilde \omega_-\over 2}$ &
${3 \sqrt 2 \over 2}[J_1+J_2]- {5 \sqrt 6 \over 4}D_z + {\cal O}(D^2/J)$
& ${3 \sqrt 2 \over 2}[J_1+J_2]+{\sqrt 2 \over 2} (2C_{x-z}+C_{y-z})
+{\cal O}(C^2/J)$ \\
$\tilde \omega_+-\tilde \omega_-$ & ${(2D_yD_z-3 \sqrt 3 D_y J_2) \over J_1}
+ {\cal O}(D^3/J^2)$ &
$C_{yz} + {\cal O} (C^2/J)$ \\
\hline \hline
\multicolumn {3} {|c|} {Neglecting Canting} \\ \hline
& DM Model & CF Model \\ \hline
$\tilde \omega_0$ &
$\left[6J_1(3J_2-\sqrt 3 D_z) -18\sqrt 3 J_2D_z+ 18D_z^2 \right]^{1/2}$ &
$\left[ 6J_1(3J_2+C_{y-z}) + 4C_{x-z}C_{y-z} + 12 J_2C_{x-z} \right]^{1/2}$ \\
${\tilde \omega_+ +\tilde \omega_-\over 2}$ & 
${3 \sqrt 2 \over 2}[J_1+J_2]- {5 \sqrt 6 \over 4}D_z + {\cal O}(D^2/J)$ &
${3 \sqrt 2 \over 2}[J_1+J_2] + {\sqrt 2 \over 2} (2C_{x-z}+C_{y-z})
+ {\cal O}(C^2/J)$ \\
$\tilde \omega_+ -\tilde \omega_-$ & $\sqrt 3 D_y$ & $0$  \\
\hline \hline
\end{tabular}
\end{table*}

Thus the SWDM is of the form
\begin{small}
\begin{eqnarray}
&& \left[ \begin{array} { c | c | c | c | c | c} \hline
a_{11} & a_{12} & -a_{12}^* & -b_{11} & -b_{12} & b_{12} \\ \hline
a_{12}^* & a_{11} & -a_{12} & -b_{12} & -b_{11} & b_{12} \\ \hline
-a_{12} & -a_{12}^* & a_{11} & b_{12} & b_{12} & -b_{11} \\ \hline
b_{11} & b_{12} & -b_{12} & -a_{11} & -a_{12}^* & a_{12} \\ \hline
b_{12} & b_{11} & -b_{12} & -a_{12} & -a_{11} & a_{12}^* \\ \hline
-b_{12}  & -b_{12} & b_{11} & a_{12}^* & a_{12} & -a_{11} \\ \hline
\end{array} \right] \ , 
\end{eqnarray}
\end{small}
where, to the same accuracy as before
\begin{eqnarray}
a_{11}&=&A_0 + 2J_2 \ , \nonumber \\ &=&
2(J_1+J_2) - 2 \sqrt 3 D_z + \Delta \nonumber \\ && \
+ 2D_y^2/J_1 + 4C_{yz}^2/(3J_1) \ ,
\end{eqnarray}
\begin{eqnarray}
a_{12} &=& -[A_1-i\alpha_1]/2 + J_2/2 - i \alpha_2 \ , \nonumber \\ &=&
-{1 \over 4} J_1 + {1 \over 2} J_2 - {\sqrt 3 \over 4} D_z
- {D_y^2 \over 4J_1} \nonumber \\ && \ - C_{yz}^2 /(12J_1) 
- {3i \over 2} {D_y J_2 \over J_1} \nonumber \\ && \
+  i \sqrt 3 [ C_{yz} + 2(D_yD_z/J_1)]/6 \nonumber \\
&\equiv& a_{12}^\prime + i a_{12}^{\prime \prime}
\end{eqnarray}
\begin{eqnarray}
b_{11} &=& B_0 = C_{xx} - C_{yy} -2 C_{yz}^2/(3J_1) \ ,
\end{eqnarray}
and
\begin{eqnarray}
b_{12} &=& -B_1/2 - (3/2)J_2 \ , \nonumber \\ &=&
{3 \over 4} J_1 - {3 \over 2} J_2 - {\sqrt 3 \over 4} D_z + {D_y^2 \over 4J_1} 
\nonumber \\ && \ + C_{yz}^2 /(12J_1) \ ,
\end{eqnarray}
where we have neglected symmetric anisotropic exchange and only kept
terms that seem to be most relevant.

Now we diagonalize the submatrices ${\bf a}$, ${\bf a}^*$, and ${\bf b}$.
The eigenvectors of these 3 $\times$ 3 matrices are
\begin{eqnarray}
\phi_\lambda  &=& {1 \over \sqrt 3} \left[ \begin{array} {c}
\lambda \\ 1  \\ - \lambda^2 \\ \end{array} \right] \ ,
\label{EIGVEC} \end{eqnarray}
where $\lambda^3=1$, so that $\lambda_1 = 1$, $\lambda_2 = (-1 + i \sqrt 3)/2$,
and $\lambda_3 = (-1 - i \sqrt 3 )/2$.
One can show that the vector given in Eq. (\ref{EIGVEC}) is an eigenvector
of ${\bf a}$ with the associated eigenvalue
\begin{eqnarray}
a(\lambda) = a_{11} + \lambda^2 a_{12} + \lambda a_{12}^* \ .
\end{eqnarray}
and is simultaneously an eigenvector of ${\bf b}$ with eigenvalue
\begin{eqnarray}
b(\lambda) = b_{11} + \lambda^2 b_{12} + \lambda b_{12} \ .
\end{eqnarray}
The three spin-wave energies are given by
\begin{eqnarray}
\tilde \omega^2 &=& \left[ \left( {a(\lambda) + a(\lambda^*) \over 2} \right)^2
-  b(\lambda)^2 \right]^{1/2} \nonumber \\ && \
+ \left( {a(\lambda) - a(\lambda^*) \over 2} \right) \ .
\end{eqnarray}
so that
\begin{eqnarray}
\tilde \omega_0 &=& [(a_{11} + 2 a_{12}^\prime )^2
- (b_{11} + 2 b_{12})^2]^{1/2} \ , \nonumber \\
\tilde \omega_{\pm} &=& 
[(a_{11} - a_{12}^\prime )^2 - (b_{11} - b_{12})^2]^{1/2} 
\pm \sqrt 3 a_{12}^{\prime \prime} \ .
\end{eqnarray}

\noindent
We thus find the spin-wave energies to be those given in Table
\ref{OMEGAX}.

\subsubsection{The Y Point}

The Y point is at $q_y=2 \pi /(\sqrt 3 a)$, $q_x=q_z=0$, so that
$\gamma_{1,3}^{(n)}=0$,
$\gamma_{2,3}^{(n)}=0$, $\gamma_{1,2}^{(1)}=1$, and
$\gamma_{1,2}^{(2)}=-1$.  Therefore the SWDM at the Y point is of the form
\begin{eqnarray}
\left[ \begin{array} { c | c | c | c | c | c} \hline
a_{11} & a_{12} & 0 & -b_{11} & -b_{12} & 0 \\ \hline
a_{12}^* & a_{11} & 0 & -b_{12} & -b_{11} & 0 \\ \hline
0 & 0 & a_{11} & 0 & 0 & -b_{11} \\ \hline
b_{11} & b_{12} & 0 & -a_{11} & - a_{12}^* & 0 \\ \hline
b_{12} & b_{11} & 0 & -a_{12} & - a_{11} & 0 \\ \hline
0 & 0 & b_{11}  & 0 & 0 & -a_{11} \\ \hline \end{array} \right] \  ,
\end{eqnarray}

\begin{table*}
\caption{\label{OMEGAY}Spin wave energies $\tilde \omega \equiv \omega/S$ at
the Y point for the DM and CF models.  The results neglecting canting are
discussed in Sec. IVE.}

\vspace{0.2 in}
\begin{tabular} { | c | c | c ||}
\hline \hline
\multicolumn {3} {|c|}  {With Canting} \\ \hline
& DM Model & CF Model \\ \hline
$\tilde \omega_-$ &
$\left[12J_1J_2- 6\sqrt 3 J_1 D_z +4J_2^2 - 14 \sqrt 3 J_2 D_z \right.  $ &
$\left[ 12J_1J_2+6J_1C_{y-z} + 4C_{x-z}C_{y-z} + 8J_2 C_{x-z} \right.$ \\
& $\left. + 18 D_z^2 + 3D_y^2\right]^{1/2} + {\cal O} (D^{5/2}/J^{3/2})$ &
$\left. + 2 J_2 C_{y-z} + 4J_2^2 + 7C_{yz}^2 \right]^{1/2}
+ {\cal O}(C^{5/2}/J^{3/2})$ \\ 
${\tilde \omega_+ + \tilde \omega_0 \over 2}$ &
$2J_1+ {5 \over 2} J_2- {7 \sqrt 3 \over 4}D_z + {\cal O}(D^2/J)$
& $2 J_1+ {5 \over 2}J_2+{1 \over 4} (6C_{x-z}+3C_{y-z})
+{\cal O}(C^2/J)$ \\
$\tilde \omega_+-\tilde \omega_0$ & $J_2 + {\sqrt 3 \over 2} D_z
+ {\cal O}(D^2/J)$ &
$J_2 + C_{x-z} - {1 \over 2} C_{y-z} + {\cal O} (C^2/J)$ \\
\hline \hline
\multicolumn {3} {|c|} {Neglecting Canting} \\ \hline
& DM Model & CF Model \\ \hline
$\tilde \omega_-$ &
$\left[12J_1J_2 -6 \sqrt 3 J_1 D_z + 4J_2^2 - 14 \sqrt 3 J_2D_z
+ 18 D_z^2 \right]^{1/2}$ &
$\left[ 12J_1J_2+6J_1C_{y-z} + 4C_{x-z}C_{y-z} \right.$ \\
& & $\left. + 8J_2 C_{x-z} + 2J_2C_{y-z} + 4J_2^2 \right]^{1/2}$ \\
${\tilde \omega_+ + \tilde \omega_0\over 2}$ &
$2J_1+ {5 \over 2} J_2- {7 \sqrt 3 \over 4}D_z + {\cal O}(D^2/J)$
& $2 J_1+ {5 \over 2}J_2+{1 \over 4} (6C_{x-z}+3C_{y-z}) +{\cal O}(C^2/J)$ \\
$\tilde \omega_+-\tilde \omega_0$ & $J_2 + {\sqrt 3 \over 2} D_z
+ {\cal O}(D^2/J)$ &
$J_2 + C_{x-z} - {1 \over 2} C_{y-z} + {\cal O} (C^2/J)$ \\
\hline \hline
\end{tabular}
\end{table*}

where
\begin{eqnarray}
a_{11}&=& A_0 + 2 J_2 \ , \nonumber \\ &=& 
2(J_1+J_2) - 2 \sqrt 3 D_z + \Delta \nonumber \\ && \
+ 2D_y^2/J_1 + 4C_{yz}^2/(3J_1) \ ,
\end{eqnarray}
\begin{eqnarray}
a_{12} &=& [A_1-i\alpha_1] - J_2/2 + i \alpha_2 \ , \nonumber \\ &=&
(J_1 - J_2)/2 + {\sqrt 3 \over 2} D_z
+ {D_y^2 \over 2J_1} - {C_{yz}^2 \over 6J_1} \nonumber \\ && \
-  i \sqrt 3 [ C_{yz} + 2(D_yD_z/J_1) - 2D_yJ_2/J_1 ]/3 \nonumber \\
&\equiv& a_{12}^\prime + i a_{12}^{\prime \prime}
\end{eqnarray}
\begin{eqnarray}
b_{11}=B_0 = C_{xx}-C_{yy}-2C_{yz}^2/(3J_1) \ ,
\end{eqnarray}
and
\begin{eqnarray}
b_{12} &=& B_1 + {3 \over 2} J_2 \nonumber \\ &=&
{3 \over 2} (J_2-J_1) + {\sqrt 3 \over 2} D_z
+ {C_{yz}^2 - 3D_y^2 \over 6J_1} \ .
\end{eqnarray}

One root for $(\tilde \omega)^2$ 
(coming from rows and columns \#3 and \#6) is
\begin{eqnarray}
{\tilde \omega_0}^2 = a_{11}^2 - b_{11}^2
\end{eqnarray}
and the remaining eigenvalues come from the projection of
the SWDM, ${\bf M}$, into the remaining subspace:
\begin{eqnarray}
{\bf M} &=& 
\left[ \begin{array} { c | c | c | c} \hline
a_{11} & a_{12} & -b_{11} & -b_{12} \\ \hline
a_{12}^* & a_{11} & -b_{12} & -b_{11} \\ \hline
b_{11} & b_{12} & -a_{11} & - a_{12}^* \\ \hline
b_{12} & b_{11} & -a_{12} & - a_{11} \\ \hline
\end{array} \right] \ . 
\label{REMAIN} \end{eqnarray}
As usual, in our analytic work, we work to second order in the
anisotropies, in the absence of which
$a_{12} = a_{12}^\prime + i a_{12}^{\prime \prime}$
is real.  In that case we have the unperturbed mode energies
\begin{eqnarray}
(\tilde \omega_{\pm}^{(0)})^2 &=& (a_{11} \pm a_{12}^\prime)^2
- (b_{11} \pm b_{12})^2 \ .
\end{eqnarray}
The perturbed (i. e. for $a_{12}^{\prime\prime} \not= 0$) secular equation is
\begin{eqnarray}
0 &=& [\tilde \omega^2 - (\tilde \omega_-^{(0)})^2]
[\tilde \omega^2 - (\tilde \omega_+^{(0)})^2] + (a_{12}^{\prime \prime})^4
\nonumber \\ && \
-2(a_{12}^{\prime \prime})^2[ \tilde \omega^2 + a_{11}^2 - (a_{12}^\prime)^2
-b_{11}^2 + b_{12}^2] \ .
\end{eqnarray}
Thereby we find the approximate solutions
\begin{eqnarray}
\tilde \omega_{\pm} &=& \tilde \omega_{\pm}^{(0)} \pm {(a_{12}^{\prime \prime}
)^2 \over \tilde \omega_\pm^{(0)} [ (\tilde \omega_+^{(0)})^2
- (\tilde \omega_-^{(0)})^2] } \nonumber \\ && \ \times
\biggl[ (\tilde \omega_\pm^{(0)})^2 + a_{11}^2 - (a_{12}^\prime)^2 - b_{11}^2 +
b_{12}^2 \biggr] \ .
\end{eqnarray}
In Table \ref{OMEGAY} we give analytical results correct only to
leading order in the anisotropies, in which case $\tilde\omega_\pm 
=\tilde\omega_\pm^{(0)}$.

\section{FIT TO THE EXPERIMENTAL SPECTRUM}

\begin{figure}
\vspace{0.5 in}
\begin{center}
\includegraphics[width=7cm]{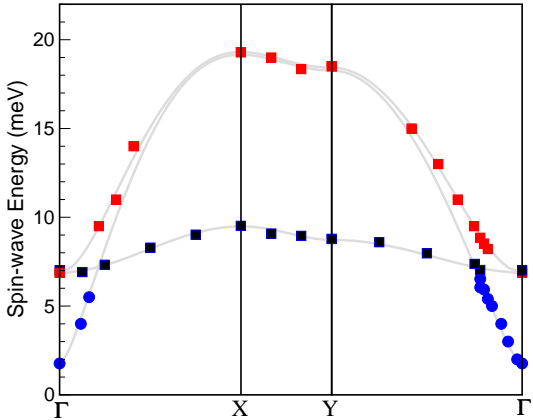} 
\caption{(Color online) Experimental spin-wave spectrum in FeJ
obtained from inelastic neutron scattering measurements.\cite{YLEE0}
The gray-lines represent the best fit which is discussed in the text
in detail. }
\label{spinwave_exp}
\end{center}
\end{figure}

Recently Matan {\it et al.}\cite{YLEE0} have measured spin-wave
spectrum in FeJ using inelastic neutron scattering and here we discuss
the comparison of our calculations with their results. The experimental
spectrum\cite{YLEE0} shown in Fig.~\ref{spinwave_exp} clearly indicates
three spin-bands. One of these bands is quite dispersionless and it is
therefore identified  to  be the ``zero-energy" mode, which is lifted
to about 8 meV due to anisotropic magnetic interactions which we
wish to identify.  In addition to this
``lifted zero-energy mode", the spectrum shows two additional gaps
at 1.7 meV (an in-plane mode) and around 7 meV (an out-of-plane
mode). The spin-wave spectrum extends up to 20 meV at the X-point. 
It is of interest to develop a model Hamiltonian which reproduces the
observed spectrum. In particular, we will see that the observed
energy splittings at the various high symmetry points provide
sensitive constraints on the interaction parameters.

In order to identify magnetic interactions in FeJ, 
in this section, we will present numerical calculations of the
spin-wave spectrum for a large number of models and compare the
results with the experimental data shown in Fig. ~\ref{spinwave_exp}.
For a quantitative comparison we define an error-factor  $R$ (i.e.
the goodness-of-fit) as follows:
\begin{equation}
R= \frac{100}{N_{\rm data}}
\sum_{i} | \omega_{\rm exp}(q_i)-\omega_{\rm cal}(q_i) |/\omega_{\rm exp}(q_i)
\end{equation}
where $N_{\rm data}$ is the number of experimental data points,
and $\omega_{\rm exp} (q_i)$ and $\omega_{\rm cal} (q_i)$ are
respectively the experimental and calculated spin wave energies
at wave-vector $q_i$. One should recognize that $R$ describes the 
percentage error 
averaged over the entire spectrum.  It is important to reproduce
the splittings at the high symmetry points even though this only
weakly affects the value of $R$. 
 
\begin{figure}
\begin{center}
\includegraphics[width=7cm]{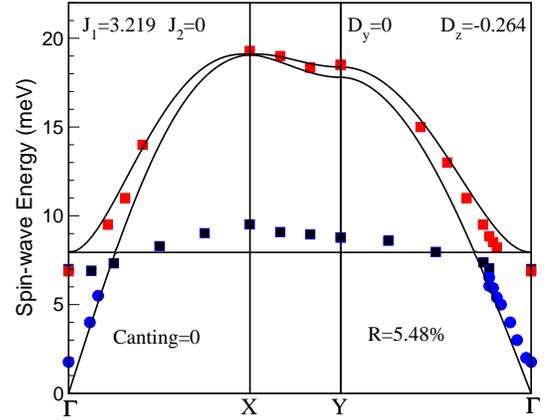} 
\caption{(Color online) Spin-wave spectrum from a simple two-parameter
$J-D_z$ model.  In these and succeeding figures the values of the
parameters of the fit are given in meV.}
\label{j1dz}
\end{center}
\end{figure} 
 
We start by comparing the experimental spectrum to a simple model
where we consider only the isotropic nearest neighbor interaction
$J_1$ and a non-zero $D_z$ in DM interaction vector (i.e. $D_x=D_y=0$).
The best fit from this simple two-parameter model is shown in
Fig.~\ref{j1dz}. The DM  interaction with only the $D_z$ term lifts the
zero-energy mode to the dispersionless energy $S\sqrt{-6\sqrt 3 JD_z}$,
as observed experimentally. However this term alone can not give the
experimentally observed small dispersion of this mode. Without $D_y$,
we do not get any canting of spins either. The effect of the $D_z$ term
is almost identical to an easy plane anisotropy. The
``lifted zero-energy mode" is found to be degenerate with the other
spin gap ({\it i.e.} the out-of-plane spin-gap) at the $\Gamma$ point. 
However, we do not get the in-plane gap as experimentally observed
at 1.7 meV. (Analytic results are given in Table \ref{OMEGA}.)

\begin{figure}
\begin{center}
\includegraphics[width=7cm]{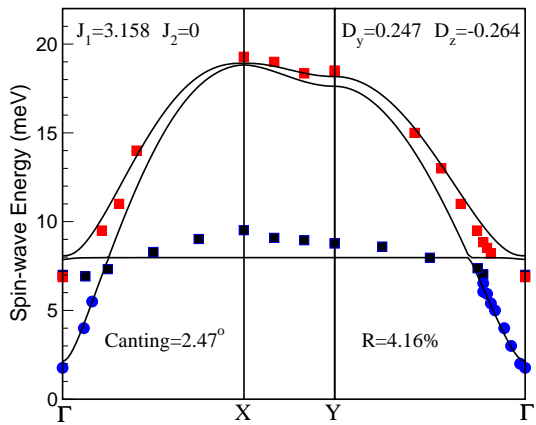} 
\caption{(Color online) Spin-wave spectrum from a simple three-parameter 
$J_1-D_y-D_z $ model. }
\label{j1dydz}
\end{center}
\end{figure}

The agreement between calculations and experiment can be improved
significantly by turning on $D_y$ in the DM-vector as 
shown in Fig.~\ref{j1dydz}. The main effect of $D_y$ term is
to  create a small in-plane anisotropy and therefore it can be
adjusted to give an in-plane gap of 1.7 meV, as experimentally observed.
This model
gives three gaps at $\Gamma$ point, for which the analytic results are
given in Table \ref{OMEGA}.  From these analytic results, one sees
that the in-plane gap is proportional to $|D_y|$ and the small splitting
between the flat-mode and the out-of-plane gap is equal to
$10D_yD_z/J$, which is quite small.

Although the experiment does not give clear results for the splittings of
the high-energy modes at the X and Y points, we see from Table \ref{OMEGAX}
that the splitting, $\delta$, at the X point is expected to be quite small
(it is second order in the perturbations). Table \ref{OMEGAY} indicates that
the splitting of the high frequency modes at the Y point is expected to
be larger than at the X point.

The $D_y$ term also causes the spins to be canted with a canting angle of
about  2.5$^{\rm o}$. However  despite a significant improvement with the
addition of $D_y$ term, we still do not get any dispersion to the flat
mode in contradiction to the experimental finding.  We also studied the
single ion term $D$ and $E$ as given in Eq.  (\ref{DE_relation})
with only $J_1$ and obtained similar results. With only DM or
single ion anisotropy without further neighbor interactions, it was
not possible to obtain the small dispersion of the flat mode
as experimentally observed. 

\begin{figure}
\begin{center}
\includegraphics[width=7cm]{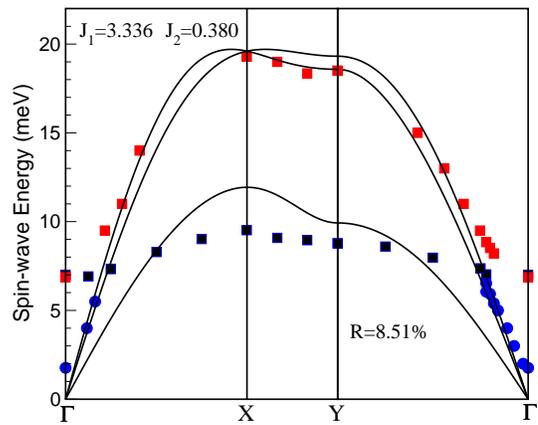} 
\caption{(Color online) Spin-wave spectrum from isotropic nn and nnn
interactions $J_1$ and $J_2$.}
\label{j1j2}
\end{center}
\end{figure}

In order to get the experimentally observed dispersion along with the
correct average energy of the nearly flat mode, one needs to consider
next-nearest neighbor interactions. In Fig. \ref{j1j2}, we show the
best fit from
a simple isotropic nn ($J_1$)  and nnn ($J_2$) interactions, which is
in good agreement with the previous work.\cite{HKB} We note
that the nnn $J_2$ interaction gives the zero-energy mode a 
significant dispersion. However in order to correctly reproduce
the flat-mode energy requires a $J_2$ which gives too strong a dispersion.
Since the Hamiltonian with isotropic $J_1$ and $J_2$ is rotationally
invariant, we do not get any gap at the $\Gamma$ point, in contrast to 
experimentally observed three gaps.
 
\begin{figure}
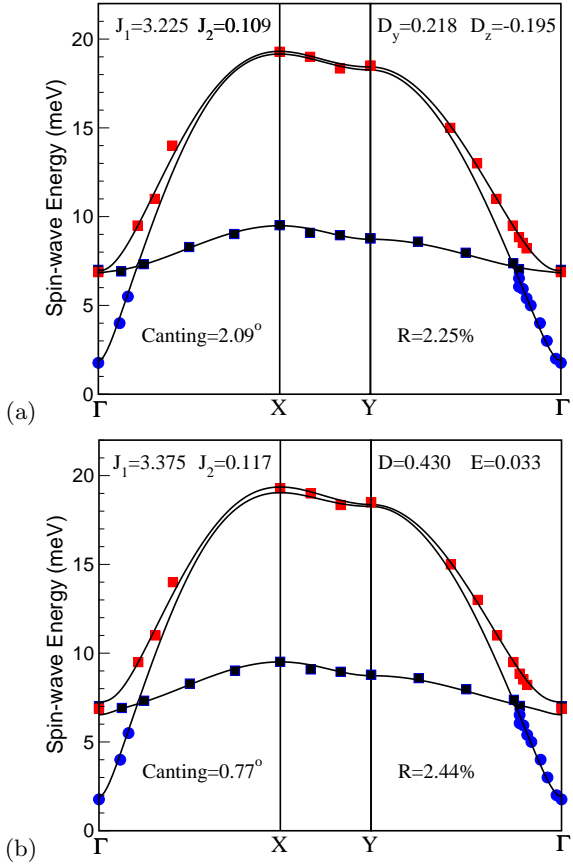

\begin{center}
(a) \includegraphics[width=7cm]{nJ1J2DyDz_fit.eps} \\
(b) \includegraphics[width=7cm]{nJ1J2DE_fit.eps} 
\caption{(Color online) Spin-wave spectrum from isotropic nn and nnn
interactions $J_1$ and $J_2$ and (a) DM-vector $(0,D_y,D_z)$ and 
(b) single ion anisotropy terms $D$ and $E$. }
\label{j1j2dydz}
\end{center}
\end{figure}

From above discussion, it is clear that one needs both a
nnn $J_2$ interaction and either a DM or a single ion anisotropy.
In Fig. ~\ref{j1j2dydz}, we show the best fit from a model which
has the isotropic nn $J_1$, nnn $J_2$ interactions and the 
DM vector $(0,D_y,D_z)$. The fit to the data is very good.
We reproduce not only the three gaps but also the small dispersion
of the flat-mode. The spin gaps at the $\Gamma$, X, and Y points obtained
numerically are in good agreement with the analytic results
given in Tables \ref{OMEGA}, \ref{OMEGAX}, and \ref{OMEGAY}.
The spin-canting angle is about 2.1$^{\rm o}$, in reasonable agreement
with the experimentally estimated value of 1.5$^{\rm o}$.\cite{YLEE0}

As an alternative to DM interactions, one can also try to explain the
experimental data based on a model which considers only the single ion
anisotropies $D$ and $E$ as given in Eq. (\ref{DE_relation}).\cite{JAPS}
In Fig.~\ref{j1j2dydz}b, we show the best fit from such a model which has
an isotropic nn $J_1$ and nnn $J_2$ and the $(D,E)$ single ion anisotropy 
instead of the DM interactions. The R-factor for this fit is
slightly larger than the R-factor obtained from DM interaction.  The main
differences are the larger splittings of the optical modes at the $\Gamma$
and X points and the  smaller canting angle for the CF model in
comparison to the DM model.  The canting angle from the CF model is only
0.8$^{\rm o}$, which is smaller than the experimental
value\cite{YLEE0} of 1.5$^{\rm o}$.

The biggest problem with the CF model is the large gaps at the
$\Gamma$ and X-points as shown in Fig.~\ref{j1j2dydz}b. The
experimental splitting of the modes at these points are much
smaller. Numerical results give a splitting of about 0.75 meV
at the $\Gamma$ point and 0.4 meV at the X-point. The splitting at
the Y-point is quite small. These results are in good agreement
with the analytical results given in Tables \ref{OMEGAX} and
\ref{OMEGAY}.

In conclusion, the observed spin-wave spectrum can be
explained by a simple model which has only nearest and
next-nearest isotropic interactions. The DM interaction
gives the best fit to explain the observed gaps and the
spin canting.
Even though the single ion terms  also fit the data well, the
quality of the fit is not as good  as the one from DM term
at the $\Gamma$ and X points.  For the Fe ion, the 
single ion anisotropy is expected to be smaller than the DM
term (because the single-ion term is second order in the spin-orbit
coupling while the DM term appears in first order). For
these reasons, we think that DM term is a natural source of
anisotropy in FeJ compound, which not only stabilizes the
observed experimental spin structure but also gives the
right spin-wave spectrum with the proper gaps and dispersion.

\subsection{Effect of Spin-Canting on the Spin-wave Spectrum}

Here we give results when the transformation to the canted
structure is omitted, i.e. $\theta$ is forced to be zero.  
One purpose of giving these results
is to show the importance of making this transformation
when calculating the gaps in the spectrum which depend
on small perturbations.  Also, we could check our calculations
by comparing these results to those of Nishiyama et al\cite{JAPS}
who omitted this transformation.  When one distorts a collinear
spin structure by the application of a small transverse field,
the effects on the spectrum are second order in the distortion
angle.  Here, however, the situation is different because
the spins before distortion are noncollinear.  As a result of this
noncollinearity the matrix element $d_y^{(1)}$ in Eq. (\ref{EFFSW})
(which would be zero for a collinear system) is of order $J$.
Thus the matrix element $\overline d_z^{(n)}$ in Eq.(\ref{TILDE})
has a contribution, $sd_y^{(n)}$, induced by canting of the
same order as that, $cd_z^{(n)}$, which existed before canting.
This fact indicates that canting should not be ignored in the
calculation of the spin-wave spectrum.

To obtain results when canting is ignored we replace quantities with
overbars in Eq. (\ref{TILDE}) by the corresponding quantities without
overbars.

\subsubsection{Zero wavevector}

Here (where $J$ denotes $J_1+J_2$)
\begin{eqnarray}
A_0&=& 4E_z+\Delta = 2J - 2 \sqrt 3 D_z + \Delta \ , 
\end{eqnarray}
\begin{eqnarray}
A_1&=& - E_x -E_y = J/2 + \sqrt 3 D_z/2 \ ,
\end{eqnarray}
\begin{eqnarray}
\alpha_1 &=& 2 d_z{(1)}=-D_y \ , \ \ \ \alpha_2=0 \ ,
\end{eqnarray}
\begin{eqnarray}
B_0&=& C_{xx}-C_{yy} \ ,
\end{eqnarray}
\begin{eqnarray}
B_1&=& E_x-E_y = -3J/2 + \sqrt 3 D_z/2 \ .
\end{eqnarray}
Then, when the anisotropy in $J_n$ is neglected, we find the
frequencies listed in Table \ref{OMEGA} under the heading
"Neglecting Canting."

\subsubsection{The X point}

The treatment including canting is now modified so that
\begin{eqnarray}
a_{11} &=& 2(J_1+J_2) - 2 \sqrt 3 D_z + \Delta \ , \nonumber \\
a_{12} &=& -[A_1-i \alpha_1]/2 + J_2/2 - i \alpha_2  \nonumber \\
&=& -J_1/4 + J_2/2 - \sqrt 3 D_z/4 -i D_y/2 \ ,\nonumber \\
b_{11} &=& C_{xx}-C_{yy} \ , \nonumber \\
b_{12}&=& -B_1/2 - 3J_2/2 \nonumber \\
&=& 3J_1/4 -3J_2/2 - \sqrt 3 D_z/4 \ .
\end{eqnarray}
Then we find the spin-wave frequencies as listed in Table \ref{OMEGAX}.
   
\subsubsection{The Y point}

Now we have
\begin{eqnarray}
a_{11}&=&A_{11}({\bf q}=0) = 2(J_1+J_2) - 2 \sqrt 3 D_z + \Delta \ ,\nonumber \\
a_{12}&=& A_{12}({\bf q}=0) = (J_1-J_2)/2 + \sqrt 3 D_z/2 + i D_y \ ,\nonumber \\
b_{11} &=& C_{xx}-C_{yy} \ , \nonumber \\
b_{12} &=& -3(J_1-J_2)/2 + \sqrt 3 D_z /2 \ ,
\end{eqnarray}
so that the frequencies are those given in Table \ref{OMEGAY}.
Unless the only anisotropy is due to $D_y$, these results are the same
as those given in Table \ref{OMEGAY} with canting, so that, except for
this case,  the effect of canting at the Y point is very small.

\subsubsection{Numerical Results and Discussion}

\begin{figure}
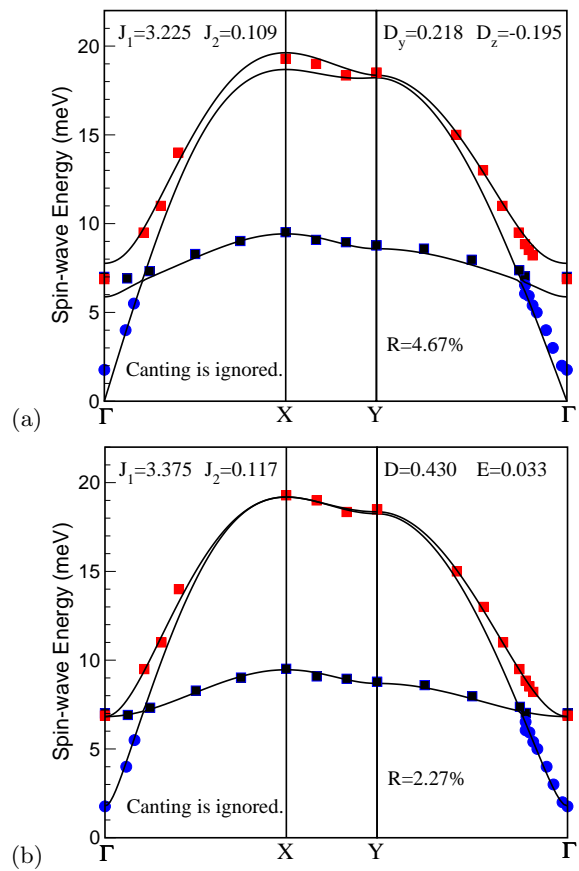

\begin{center}
(a) \includegraphics[width=7cm]{nj1j2dydz_nocanting.eps} \\
(b) \includegraphics[width=7cm]{nj1j2de_nocanting.eps} 
\caption{(Color online) Spin-wave spectrum for the DM (a) and CF (b) models
when the spin-canting is ignored. We used the same parameters
as in Fig. \protect{\ref{j1j2dydz}} to emphasize the effect of spin-canting
on the spectrum.}
\label{no_canting}
\end{center}
\end{figure}

The effect of canting can be deduced from the epxressions for the
matrix elemets given in Appendix C. One sees that the rotation to the
canted spin structure induces terms of the form $D_y^2/J_1$ into the
SWDM matrix elements.  Also the quantity 
$\alpha_1 \equiv 2 \overline d_z^{(1)}$ is transformed (by the rotation
to the canted system of coordinates) from the value $D_y$
(neglecting the anisotropy in $J_n$) to $(C_{yz}+ 2 D_zD_y/J_1)\sqrt 3
- D_yJ_2/J_1$.  Suprprisingly perhaps, this rotation through an angle
of order $D_y/J_1$ also introduces a contribution to $\alpha_2$ of
order $J_2D_y/J_1$. A qualitative understanding of what is
going on here is obtained by noting that the coefficient $\overline d_z$
is associated with an interaction proportional to
$S_x^{\rm loc,f} S_y^{\rm loc,f}$,
where the components are taken in the ``final'' {\it i. e.} canted
coordinate system.  This term directly affects the spin-wave energies.
This term arises from rotating an interaction in the local uncanted
coordinate system of the form $d_y S_x^{\rm loc,i}S_z^{\rm loc,i}$,
a term which for collinear systems is zero, but for this noncollinear
system is of order $J_1$, as given in Eq. (\ref{PARAM}).
Since these quantities control several of the splittings, it is clear
that neglecting canting can have a major impact on these splittings,
as shown in Fig. \ref{no_canting} and the tables. From Fig.~\ref{j1j2dydz}(a),
it is clear that the quality of the fit is significantly degraded for the DM
term, in particular at the $\Gamma$ and $X$-points where big gaps are opened. 

The effect of neglecting canting
is opposite for the CF term. The quality of the fit is improved
because the large splitting at $\Gamma$ (see Fig.~\ref{j1j2dydz}(b)) is now
zero (Fig.~\ref{no_canting}(b)).
For instance, for zero wavevector the splitting $\tilde \omega_+ -
\tilde \omega_-$ (given by $2\sqrt 3 \alpha$) is linear in $D_y$ when
canting is neglected, whereas when the canting was taken into account
the splitting was found to be proportional to $C_{yz} + 2D_zD_y/J_1$.
To calculate this splitting it is therefore not qualitatively correct to
ignore the canting.  In addition, when canting is ignored and single-ion
anisotropy is not present, neglecting canting leads to
$\tilde \omega_0=0$, even when ${\bf D}$ is nonzero.

At the X point the effects of canting are similar.  For the low
energy mode the transformation to the canted structure introduces a
term proportional to $D_y^2$ which is only important if all the other
anisotropies are very small or vanishing.  However, as Table \ref{OMEGAX}
indicates, the splitting between $\omega_+-\omega_-$ is proportional
to $D_y$ when canting is neglected, but is proportional to
$C_{yz}+2D_yD_z/J_1 - 3 \sqrt 3 J_2D_y/J_1$ when canting is taken into account.

At the Y point the effects of canting are less important. Indeed
$\omega_0$ and $\omega_+$ are insensitive to canting and only
in $\omega_-$ does one see that canting introduces a term of order
$D_y^2$. 

In conclusion, it is important that one should not ignore the
spin-canting when spin-wave spectrum is calculated. Even though
the CF single ion term gives much better fit when the canting
is ignored (see Figs.~\ref{j1j2dydz} and \ref{no_canting}), 
it is not an appropriate approximation, as we have shown.
When the CF single ion term is properly treated, it 
actually gives a worse fit to the data and therefore it is more logical
to conclude that the dominant source of the anisotropy in FeJ
compound is the DM term.

\section{Conclusions}

We have presented a detailed study of the magnetic structure and
spin-wave spectrum in FeJ for the most generic nn and nnn 
exchange interactions including Dzyaloshinskii-Moriya and
the single ion anisotropy terms of Inami {\it et al.}\cite{INAMI}
Our conclusions can be summarized as follows.

\begin{itemize}
\item 
We first discuss
the allowed magnetic structures from a representations analysis.
The experimentally observed ``umbrella" structure of 
{\bf q}=0 with positive chirality arises from a one dimensional
representation which exhibits three-fold symmetry. 
Our analysis and discussion of broken symmetry suggests that the
characterization by irreducible representations is
more fundamental than that by chirality,
even though chirality has been widely invoked in the literature.

\item
Representation analysis reveals the existence of an $x$-$y$-like
two dimensional representation which has sixth order in-plane anisotropy.
In this representation one can have Kagom\'e layers with a net
in-plane moment which can be stacked either ferromagnetically
or antiferromagnetically.  The analysis of Ref. \onlinecite{WILLS}
did not find wavefunctions of this type.  However, the observed
magnetic ground state of a NaV-jarosite\cite{NAV} consists of such
ferromagnetic Kagom\'e planes which are coupled antiferromagnetically.

\item
We presented detailed spin-wave calculations for the most generic
spin Hamiltonian allowed by symmetry. We derived analytic expressions
for the spin-wave energies
at the $\Gamma$, $X$, and $Y$ points in the Brillouin zone.
The agreement between our analytic and numerical results
tends to support the correctness of both approaches.
Since, the spectrum is invariant under a change of sign of $D_y$,
the sign of this parameter remains undetermined by experiment.

\item
We obtained the spin-wave spectrum numerically throughout the Brillouin
zone for a large number of
models with increasing complexity and compared the results with the
recent experimental data.\cite{YLEE0} We obtained a nn exchange constant of
$J_1=3.225$ meV, which is in good agreement with previous
measurements.\cite{YLEE0}
We found that the observed small dispersion can  be explained
by a nnn isotropic interaction  of $J_2 = 0.11$. The observed gaps
at the $\Gamma$, $X$, and $Y$ points are best explained by considering only
the DM interaction with $|D_y|=0.218$ meV and $D_z=-0.195$ meV together
with nn and nnn interactions. The D$_z$ term is adjusted to
lift the zero-energy mode to about 8 meV with a small dispersion due
to $J_2$  interaction. A fit of similar quality can also be obtained using
the crystal field (CF) model of Ref. \onlinecite{JAPS} but the resulting
anisotropy is about 10\% of $J_1$, too large
a value for the S state Fe$^{3+}$ ion which is supposed to be quite isotropic. 
The calculated canting angle is 2.07$^{\rm o}$ and 0.77$^{\rm o}$
for DM and CF models, respectively, compared to the experimentally
deduced value\cite{YLEE0} of $1.5^{\rm o}$.

\item
Finally we discuss the effect of the canting on the spin-wave quantization 
axis which was neglected by Inami {\it et al.}\cite{JAPS} We found that
neglecting canting is not a good approximation for calculating the
spin-wave energy gaps at the high symmetry points in the Brillouin zone.
When the CF is treated properly, it gives larger gaps at
the $\Gamma$ point than those from DM interaction. The experiment does
not indicate significant splitting at the $\Gamma$ point and therefore
it suggests that the DM term works better to explain the data.

\end{itemize}

\noindent {\bf ACKNOWLEDGEMENT:}
We thank K. Matan and Y. S. Lee for many fruitful discussions and for
sharing their spin-wave data prior to publication.  

\appendix

\section{Boson Hamiltonian} \label{BOSON}
To construct the boson Hamiltonian it is convenient to introduce
spherical components ($+=x+iy$, $-=x-iy$) for
interactions between site $\tau$ in unit cell $i$ located at
${\bf R}_i$ and site $\tau'$ in unit cell $j$ located at ${\bf R}_j$.
If these sites are $n$th nearest neighbors, then we write
\begin{eqnarray}
\overline {\cal I}^{zz}_{\tau \tau'} (i,j) 
&=& \overline {\cal I}_{zz}^{(n)} \ , \nonumber \\  
\overline {\cal I}^{\pm \pm}_{\tau \tau'}(i,j)
&=& {1 \over 4} [ \overline {\cal I}_{xx}^{(n)}
-i \overline {\cal I}_{xy}^{(n)} -i \overline {\cal I}_{yx}^{(n)} 
- \overline {\cal I}_{yy}^{(n)}] \nonumber \\ 
\overline{\cal I}^{\pm \mp}_{\tau \tau'}(i,j)
&=& {1 \over 4} [ \overline {\cal I}_{xx}^{(n)} 
\pm i \overline {\cal I}_{xy}^{(n)} \mp \overline {\cal I}_{yx}^{(n)} 
+ \overline {\cal I}_{yy}^{(n)}] 
\end{eqnarray}
and similarly for the single-ion anisotropy matrix ${\bf C}$.  Now we
invoke the Holstein-Primakoff transformation in the form
\begin{eqnarray}
S_{i\tau}^z &=& S - a_{i \tau}^\dagger a_{i\tau} \ , \
S_{i \tau}^+ = \sqrt{2S} a_{i \tau} \ , \ 
S_{i \tau}^- = \sqrt{2S} a_{i \tau}^\dagger \ .
\end{eqnarray}
Then apart from a constant the Hamiltonian in terms of Bose operators is
\begin{eqnarray}
{\cal H} &=& 2S \sum_{i \tau}
\Biggl[ - C_\tau^{zz} a_{i \tau}^\dagger a_{i \tau}
+ C_{\tau}^{++} a_{i \tau} a_{i \tau} + C_{\tau}^{--} a_{i \tau}^\dagger
a_{i \tau}^\dagger \nonumber \\ && \
+ C_{\tau}^{+-} a_{i \tau}^\dagger a_{i \tau}
+ C_{\tau}^{-+} a_{i \tau}^\dagger a_{i \tau} \Biggr] \nonumber \\
&& \ + S {\sum_{i \tau, j \tau'}}^\prime 
\Biggl[ - {1 \over 2} \overline {\cal I}_{\tau \tau'}^{zz}(i,j)
(a_{i \tau}^\dagger a_{i \tau} + a_{j \tau'}^\dagger a_{j \tau'})
\nonumber \\ && \ + \overline{\cal I}_{\tau \tau'}^{++} (i,j)
a_{i \tau} a_{j\tau'} + \overline{\cal I}_{\tau \tau'}^{--} (i,j)
a_{i \tau}^\dagger a_{j\tau'}^\dagger
\nonumber \\ && \ 
+ \overline{\cal I}_{\tau \tau'}^{+-} (i,j) a_{i \tau} a_{j\tau'}^\dagger
+ \overline{\cal I}_{\tau \tau'}^{-+} (i,j) a_{i \tau}^\dagger a_{j\tau'}
\Biggr] \ ,
\end{eqnarray}
where the prime on the summation excludes the term with
$(i,\tau)=(j,\tau')$.  Now write
\begin{eqnarray}
a_{i\tau}&=& {1 \over \sqrt N} \sum_{\bf q}
e^{i {\bf q} \cdot ({\bf R}_i+\tauv)} 
a_\tau ({\bf q}) \ , \nonumber \\
a_{i\tau}^\dagger &=& {1 \over \sqrt N} \sum_{\bf q}
e^{-i {\bf q} \cdot ({\bf R}_i+\tauv)} a_\tau^\dagger ({\bf q}) \ ,
\end{eqnarray}
so that
\begin{eqnarray}
{\cal H} = E_0 + \sum_{\bf q} {\cal H}({\bf q}) \ ,
\end{eqnarray}
where $E_0$ is the ground state energy and
\begin{eqnarray}
{\cal H}({\bf q})/S &=& \sum_{\mu \nu} A_{\mu \nu}({\bf q})
a_\mu^\dagger ({\bf q}) a_\nu({\bf q}) \nonumber \\ && \
+ {1 \over 2} \sum_{\alpha
\beta} B_{\alpha \beta}({\bf q}) a_\alpha^\dagger ({\bf q}) a_\beta^\dagger
(-{\bf q}) \nonumber \\ && \ + {1 \over 2} \sum_{\alpha
\beta} B_{\alpha \beta}({\bf q})^* a_\alpha ({\bf q}) a_\beta (-{\bf q}) \ ,
\end{eqnarray}
where
\begin{eqnarray}
A_{\tau,\tau'}({\bf q}) &=& \Biggl( [2C_\tau^{+-} + 2C_\tau^{-+}
- 2C_\tau^{zz} \nonumber \\ &-&
\sum_{\tau''} \overline {\cal I}_{\tau,\tau''}^{zz} (q=0) ]
\delta_{\tau , \tau'}
+ \overline {\cal I}^{+-}_{\tau' \tau}({\bf q})
+ \overline {\cal I}^{-+}_{\tau \tau'}(-{\bf q}) \Biggr) \nonumber \\
&=& \Biggl( [C_{xx} + \overline C_{yy} - 2\overline C_{zz}
- \sum_{\tau''} \overline {\cal I}_{\tau,\tau''}^{zz} (q=0) ]
\delta_{\tau , \tau'}
\nonumber \\ && \ + \overline {\cal I}^{+-}_{\tau' \tau}({\bf q})
+ \overline {\cal I}^{-+}_{\tau \tau'}(-{\bf q}) \Biggr) \nonumber \\
\label{ADEF} \end{eqnarray}
\begin{eqnarray}
B_{\tau,\tau'}({\bf q}) &=& 4 C^{--}_\tau \delta_{\tau,\tau'} +
2 \overline {\cal I}_{\tau \tau'}^{--}(-{\bf q}) \nonumber \\
&=& [C_{xx} - \overline C_{yy}] \delta_{\tau,\tau'} +
2 \overline {\cal I}_{\tau \tau'}^{--}(-{\bf q}) \ ,
\label{BDEF} \end{eqnarray}
where
\begin{eqnarray}
\overline {\cal I}_{\tau,\tau'}^{\alpha , \beta} ({\bf q}) &=& \sum_i
\overline{\cal I}_{\tau,\tau'}^{\alpha , \beta} (i,j) e^{i {\bf q} \cdot (
\tauv+ {\bf R}_{ij} - \tauv')} \ ,
\label{GAMMAEQ} \end{eqnarray}
where ${\bf R}_{ij}={\bf R}_i - {\bf R}_j$ and
the superscripts assume the values $+$, $-$, and $z$.  Since each
site is a center of inversion symmetry, the exponential
factor in the sum over $i$ in Eq. (\ref{GAMMAEQ}) can be replaced by a
cosine:
\begin{eqnarray}
\overline {\cal I}_{\tau,\tau'}^{\alpha , \beta} ({\bf q}) &=& \sum_i
{\cal I}_{\tau,\tau'}^{\alpha , \beta} (i,j) \cos[ {\bf q} \cdot (
\tauv+ {\bf R}_{ij} - \tauv')]
\label{COSEQ} \end{eqnarray}
and $\overline {\cal I}({\bf q})$ is an even function of ${\bf q}$.
If ${\bf G}$ is a vector of the reciprocal lattice (such that
${\bf G} \cdot {\bf R}$ is a multiple of $2 \pi$), then
\begin{eqnarray}
A_{\tau \tau'}({\bf q} + {\bf G}) &=& A_{\tau \tau'}({\bf q})
e^{i {\bf G} \cdot (\tauv'-\tauv)}\nonumber \\ &=& A_{\tau \tau'}({\bf q})
\cos [ {\bf G} \cdot (\tauv'-\tau)] \nonumber \\
B_{\tau \tau'}({\bf q} + {\bf G}) &=& B_{\tau \tau'}({\bf q})
e^{i {\bf G} \cdot (\tauv'-\tauv)}\nonumber \\ &=& B_{\tau \tau'}({\bf q})
\cos [ {\bf G} \cdot (\tauv'-\tau)]  \ .
\label{GVEC} \end{eqnarray}
Here we used the fact that for the Kagom\'e lattice, the $\tauv$'s are
half a lattice vector, so that $e^{i{\bf G} \cdot (\tauv' - \tauv)}=
\cos [ {\bf G} \cdot (\tauv'-\tau)]= \pm 1$.

\section{NORMAL MODES} \label{MODES}

We write the normal mode operators as
\begin{eqnarray}
X^\dagger_\mu ({\bf q}) &=& \sum_\alpha [c_\alpha^\mu({\bf q})
a_\alpha^\dagger({\bf q})
+ d_\alpha^\mu ({\bf q}) a_\alpha (-{\bf q})] \ ,
\end{eqnarray}
and these are determined by
\begin{eqnarray}
[ {\cal H} , X^\dagger_\mu ({\bf q}) ] = \omega^{(\mu)} ({\bf q})
X^\dagger_\mu({\bf q}) \ .
\label{MODEEQ} \end{eqnarray}
If there are $p$ spins per unit cell, then we expect that $p$ of
the $\omega^{(\mu)}$ will be nonnegative and $p$ will be nonpositive.

Equation (\ref{MODEEQ}) gives
\begin{eqnarray}
&& \tilde \omega^{(\mu)} ({\bf q})
\sum_\alpha [c^\mu_\alpha ({\bf q}) a_\alpha^\dagger
({\bf q}) + d^\mu_\alpha ({\bf q}) a_\alpha (-{\bf q}) ] \nonumber \\
&=& \sum_{\alpha , \beta} \Biggl[ c_\alpha^\mu ({\bf q}) 
A_{\beta , \alpha}({\bf q}) 
a_\beta^\dagger ({\bf q}) + c_\alpha^\mu ({\bf q}) B_{\alpha, \beta}^*({\bf q})
a_\beta(-{\bf q}) \Biggr] \nonumber \\ && \ \  -
\sum_{\alpha , \beta} \Biggl[ d_\alpha^\mu ({\bf q})
A_{\alpha , \beta}(-{\bf q}) a_\beta(-{\bf q})
\nonumber \\ && \ + d_\alpha^\mu ({\bf q}) B_{\alpha, \beta}(-{\bf q})
a_\beta^\dagger({\bf q}) \Biggr] \ ,
\end{eqnarray}
where again $\tilde \omega \equiv \omega/S$.

\subsection{Eigenvalue Problem}

This leads to the eigenvalue problem
\begin{eqnarray}
\tilde \omega^{(\mu)} ({\bf q}) c_\alpha^\mu ({\bf q}) &=&
\sum_\beta [ A_{\alpha , \beta}({\bf q}) c_\beta^\mu ({\bf q})
- B_{\beta , \alpha} (-{\bf q}) d_\beta^\mu ({\bf q}) ] \nonumber \\ 
\tilde \omega^{(\mu)} ({\bf q}) d_\alpha^\mu ({\bf q}) &=&
\sum_\beta [ B_{\beta , \alpha}({\bf q})^* c_\beta^\mu ({\bf q})
- A_{\beta , \alpha} (-{\bf q}) d_\beta^\mu ({\bf q}) ] \ ,
\end{eqnarray}
which can be written as
\begin{small}
\begin{eqnarray}
\left[ \begin{array} {c c }
{\bf A}({\bf q}) & - \tilde {\bf B}(-{\bf q}) \\
\tilde {\bf B}({\bf q})^* & - \tilde {\bf A}(-{\bf q}) \\
\end{array} \right] \left[ \begin{array} {c} {\bf c}^\mu ({\bf q})
\\ {\bf d}^\mu ({\bf q}) \\ \end{array} \right]
&=& \tilde \omega^{(\mu)} ({\bf q})
\left[ \begin{array} {c} {\bf c}^\mu({\bf q}) \\
{\bf d}^\mu ({\bf q}) \\ \end{array} \right] \  ,
\end{eqnarray}
\end{small}
where ${\bf c}^\mu ({\bf q})$ is a column vector with components
$c^\mu_1({\bf q})$, $c^\mu_2({\bf q})$, $\dots$,
$c^\mu_p({\bf q})$, where $p=3$ is the number of spins in the unit cell.

We now note that because the Hamiltonian is Hermitian and taking account
of inversion symmetry, we have
\begin{eqnarray}
A_{\tau \tau'} ({\bf q}) &=& A_{\tau' \tau} ({\bf q})^*
= A_{\tau \tau'}(-{\bf q}) \ .
\label{EQA} \end{eqnarray}
Similarly we write
\begin{eqnarray}
B_{\tau \tau'}({\bf q}) &=& B_{\tau' \tau}(-{\bf q}) = B_{\tau \tau'}(-{\bf q}) \ .
\label{EQB} \end{eqnarray} 

Thus the eigenvalue problem is
\begin{eqnarray}
\left[ \begin{array} {c c }
{\bf A}({\bf q}) & - {\bf B}({\bf q}) \\
{\bf B}({\bf q})^* & -{\bf A}({\bf q})^* \\
\end{array} \right] \left[ \begin{array} {c} {\bf c}^\mu ({\bf q})
\\ {\bf d}^\mu ({\bf q}) \\ \end{array} \right]
&=& \tilde \omega^{(\mu)} ({\bf q})
\left[ \begin{array} {c} {\bf c}^\mu({\bf q}) \\
{\bf d}^\mu ({\bf q}) \\ \end{array} \right] \  .
\label{MEQ} \end{eqnarray}

We now show that the roots come in pairs with opposite signs.  Equation
(\ref{MEQ}) is
\begin{eqnarray}
{\bf A} {\bf c}^\mu - {\bf B} {\bf d}^\mu &=&
\tilde \omega^{(\mu)} {\bf c}^\mu \nonumber \\ 
{\bf B}^* {\bf c}^\mu - {\bf A}^* {\bf d}^\mu &=&
\tilde \omega^{(\mu)} {\bf d}^\mu \ .
\end{eqnarray}
Take the complex conjugate of these equations, change the signs of both sides
of each equations, and reorder the equations to get
\begin{eqnarray}
{\bf A} [{\bf d}^\mu]^* - {\bf B} [{\bf c}^\mu]^* &=&
- \tilde \omega^{(\mu)} [{\bf d}^\mu]^* \nonumber \\ 
{\bf B}^* [{\bf d}^\mu]^* - {\bf A}^* [{\bf c}^\mu]^* &=&
- \tilde \omega^{(\mu)} [{\bf c}^\mu]^* \ ,
\end{eqnarray}
where we used the fact that the $\tilde \omega$'s are real.  This shows that
from an eigenvector with eigenvalue $\tilde \omega^{(\mu)}$ we can construct a
related eigenvector with eigenvalue $-\tilde \omega^{(\mu)}$.  One can show
that the eigenvalues at wavevector ${\bf q}$ are identical to those at
wavevector $-{\bf q}$, as expected in view of the inversion symmetry of
the lattice. Thus we may write
\begin{eqnarray}
&& \left[ \begin{array} {c c } {\bf A}({\bf q}) & - {\bf B}({\bf q}) \\
{\bf B}({\bf q})^* & - {\bf A}({\bf q})^* \\ \end{array} \right]
\left[ \begin{array} {c c} {\bf c}({\bf q}) & {\bf d}({\bf q})^* \\
{\bf d}({\bf q}) & {\bf c}({\bf q})^* \\ \end{array} \right] \nonumber \\
&=& \left[ \begin{array} {c c } {\bf c}({\bf q}) & {\bf d}({\bf q})^* \\
{\bf d}({\bf q}) & {\bf c}({\bf q})^* \\ \end{array} \right]
\left[ \begin{array} {c c} \tilde \omegav ({\bf q}) & 0 \\ 0 & - \tilde \omegav ({\bf q}) \\
\end{array} \right] \ ,
\label{54EQ} \end{eqnarray}
where ${\bf c}({\bf q})$ and ${\bf d}({\bf q})$ are matrices whose columns
are the column vectors ${\bf c}^\mu({\bf q})$ and ${\bf d}^\mu({\bf q})$,
respectively, and
$\tilde \omegav$ is a diagonal matrix with entries
$\tilde \omega^{(\mu)}({\bf q})$.

\subsection{Dependence on Reciprocal Lattice Vector}

Recall the dependence of the matrices on the reciprocal lattice
vector as recorded in Eq. (\ref{GVEC}). Then the eigenvalue problem
at wavevector ${\bf q} + {\bf G}$ may be written in the form
\begin{eqnarray}
&& \tilde \omega^{(\mu)} ({\bf q}+{\bf G}) c_\alpha^\mu ({\bf q}+{\bf G}) =
\sum_\beta [ A_{\alpha , \beta}({\bf q}) c_\beta^\mu ({\bf q}+{\bf G})
\nonumber \\ && \
- B_{\beta , \alpha} (-{\bf q}) d_\beta^\mu ({\bf q}+{\bf G}) ]
e^{i{\bf G} \cdot (\tauv_\beta - \tauv_\alpha)} \nonumber \\ 
&& \tilde \omega^{(\mu)} ({\bf q}+{\bf G}) d_\alpha^\mu ({\bf q}+{\bf G}) =
\sum_\beta [ B_{\beta , \alpha}({\bf q})^* c_\beta^\mu ({\bf q}+{\bf G})
\nonumber \\ && \
- A_{\beta , \alpha} (-{\bf q}) d_\beta^\mu ({\bf q}+{\bf G}) ]
e^{i{\bf G} \cdot (\tauv_\beta - \tauv_\alpha)} \ .
\end{eqnarray}
Set $\tilde \omega^{(\mu)}({\bf q}+{\bf G})=\tilde \omega^{(\mu)} ({\bf q})$
and
\begin{eqnarray}
c^\mu_\beta({\bf q}+ {\bf G}) &=& c^\mu_\beta ({\bf q})
e^{-i{\bf G} \cdot \tauv_\beta} \nonumber \\
d^\mu_\beta({\bf q}+ {\bf G}) &=& d^\mu_\beta ({\bf q})
e^{-i{\bf G} \cdot \tauv_\beta} \ .
\label{GTRANS} \end{eqnarray}
Then we recover the previous equations.  So we conclude that under
addition of ${\bf G}$, the eigenvalues remain invariant, but the
wavefunctions change according to Eq. (\ref{GTRANS}).

\subsection{Orthogonality}

Taking the Hermitian conjugate of eigenvalue equations, we get
\begin{eqnarray}
{\bf c}^\mu({\bf q})^\dagger {\bf A}({\bf q})
- {\bf d}^\mu ({\bf q})^\dagger {\bf B}({\bf q})^\dagger
&=& \tilde \omega^{(\mu)} ({\bf q}) {\bf c}^\mu({\bf q})^\dagger \nonumber \\
{\bf c}^\mu({\bf q})^\dagger \tilde {\bf B}({\bf q})
- {\bf d}^\mu ({\bf q})^\dagger \tilde {\bf A}({\bf q})
&=& \tilde \omega^{(\mu)} ({\bf q}) {\bf d}^\mu({\bf q})^\dagger \ .
\end{eqnarray}
so that
\begin{eqnarray}
&& \left[ \begin{array} {c } 
{\bf c}^\mu({\bf q})^\dagger \ , - {\bf d}^\mu({\bf q})^\dagger
\end{array} \right] \left[ \begin{array} {c c }
{\bf A}({\bf q}) & - {\bf B}({\bf q}) \\
{\bf B}({\bf q})^* & -{\bf A} ({\bf q})^* \\
\end{array} \right] \nonumber \\ && \  = \tilde \omega^{(\mu)} ({\bf q})
\left[ \begin{array} {c } {\bf c}^\mu({\bf q})^\dagger \ ,
- {\bf d}^\mu ({\bf q})^\dagger \end{array} \right] \  .
\end{eqnarray}
Thus from any right solution we have constructed the associated left solution.
Now consider
\begin{eqnarray}
W_{\mu \nu} &\equiv & \left[ c^\mu ({\bf q})^\dagger \ ,
- d^\mu ({\bf q})^\dagger \right] \left[ \begin{array} {c c }
{\bf A}({\bf q}) & - {\bf B}({\bf q}) \\
{\bf B}({\bf q})^* & -{\bf A} ({\bf q})^* \\
\end{array} \right] \nonumber \\ && \ \times
\left[ \begin{array} {c} c^\nu ({\bf q}) \\
d^\nu ({\bf q}) \\ \end{array} \right] \ .
\end{eqnarray}
Using the fact that the left and right vectors are eigenvectors, we have
\begin{eqnarray}
W_{\mu \nu} &=& \tilde \omega^{(\mu)} ({\bf q})^*
\left[ \begin{array} {c } 
{\bf c}^\mu({\bf q})^\dagger \ , - {\bf d}^\mu({\bf q})^\dagger
\end{array} \right] \left[ \begin{array} {c} c^\nu ({\bf q}) \\
d^\nu ({\bf q}) \\ \end{array} \right] \nonumber \\ &=&
\tilde \omega^{(\nu)} ({\bf q}) \left[ \begin{array} {c } 
{\bf c}^\mu({\bf q})^\dagger \ , - {\bf d}^\mu({\bf q})^\dagger
\end{array} \right] \left[ \begin{array} {c} c^\nu ({\bf q}) \\
d^\nu ({\bf q}) \\ \end{array} \right] \ .
\end{eqnarray}
If $\mu=\nu$ this shows that $\tilde \omega^{(\mu)} ({\bf q})$ is real.  Also
\begin{eqnarray}
\left[ \begin{array} {c } 
{\bf c}^\mu({\bf q})^\dagger \ , - {\bf d}^\mu({\bf q})^\dagger
\end{array} \right] \left[ \begin{array} {c} c^\nu ({\bf q}) \\
d^\nu ({\bf q}) \\ \end{array} \right] = 0 , \ \ \ \mu \not= \nu \ .
\end{eqnarray}
(If we have degenerate eigenvalues, we can choose the wavefunctions
to preserve this orthogonality.)

A more general orthogonality is 
\begin{eqnarray}
\left[ \begin{array} { c c } {\bf c}^\dagger ({\bf q}) & - {\bf d}^\dagger
({\bf q}) \\ - \tilde {\bf d}({\bf q}) & \tilde {\bf c} ({\bf q}) \\ 
\end{array} \right] 
\left[ \begin{array} { c c } {\bf c}({\bf q}) & {\bf d}({\bf q})^* \\
{\bf d}({\bf q}) & {\bf c}({\bf q})^* \\ \end{array} \right] & =&
{\cal I} \ ,
\end{eqnarray}
where $\cal I$ is the $2p \times 2p$ unit matrix.  

\subsection{Summary}

We have the transformation to normal modes as
\begin{eqnarray}
X^\dagger_\mu ({\bf q}) &=& \sum_\alpha [c_\alpha^\mu({\bf q})
a_\alpha^\dagger({\bf q})
+ d_\alpha^\mu ({\bf q}) a_\alpha (-{\bf q})] \ ,
\label{NEGO} \end{eqnarray}
and also
\begin{eqnarray}
X_\mu ({\bf q}) &=& \sum_\alpha [c_\alpha^\mu({\bf q})^*
a_\alpha({\bf q}) + d_\alpha^\mu ({\bf q})^* a_\alpha^\dagger (-{\bf q})] \ ,
\end{eqnarray}
where $\mu$ is restricted so that $\tilde \omega^{(\mu)}>0$.
We require that $X_\mu({\bf q})$ obey Bose commutation relations:
\begin{eqnarray}
[X_\mu ({\bf q}) , X_\nu^\dagger ({\bf q}) ] = \delta_{\mu,\nu} \ .
\end{eqnarray}
This gives
\begin{eqnarray}
\sum_\tau \left( c^\mu_\tau({\bf q})^* c_\tau^\nu ({\bf q})
- d^\mu_\tau({\bf q})^* d_\tau^\nu ({\bf q}) \right)
&=& \delta_{\mu , \nu} \ .
\end{eqnarray}
The inverse transformation is
\begin{eqnarray}
a_\alpha^\dagger ({\bf q}) &=& \sum_\mu [c_\alpha^\mu({\bf q})^*
X_\mu({\bf q})^\dagger - d_\alpha^\mu ({\bf q}) X_\mu (-{\bf q})] \ ,
\nonumber \\
a_\alpha ({\bf q}) &=& \sum_\mu [c_\alpha^\mu({\bf q})
X_\mu({\bf q}) - d_\alpha^\mu ({\bf q})^* X_\mu (-{\bf q})^\dagger] \ .
\end{eqnarray}

We record here the dependence on reciprocal lattice vector:
\begin{eqnarray}
a_\alpha^\dagger ({\bf q}+{\bf G}) &=& \sum_\mu [c_\alpha^\mu({\bf q})^*
X_\mu({\bf q})^\dagger - d_\alpha^\mu ({\bf q}) X_\mu (-{\bf q})]
e^{i{\bf G} \cdot \tauv_\alpha} \ ,
\nonumber \\
a_\alpha ({\bf q}+{\bf G}) &=& \sum_\mu [c_\alpha^\mu({\bf q})
X_\mu({\bf q}) - d_\alpha^\mu ({\bf q})^* X_\mu (-{\bf q})^\dagger]
e^{-i{\bf G} \cdot \tauv_\alpha} \ .
\end{eqnarray}
To use this we would assume that ${\bf q}$ is in the first Brillouin zone
and ${\bf G}$ is the vector needed to bring the actual wavevector back
into the first Brillouin zone.

\section{EXPLICIT EVALUATION OF MATRIX ELEMENTS} \label{EVALUATE}

We now develop explicit expressions for the constants appearing in these
matrices.  We use Eq.  (\ref{TILDE}) to relate the overlined
coefficients to the bare coefficients, which are defined in Eqs.
(\ref{CDEF}) and (\ref{PARAM}).
We will work to second order in the perturbations from
isotropic exchange. To this order
\begin{eqnarray}
\sin(2\theta ) &=&  {H_y \over H} =  {H_y \over H_z}
\end{eqnarray}
and
\begin{eqnarray}
\cos (2 \theta) &=& {H_z \over \sqrt{H_z^2+H_y^2} } =
1 - {H_y^2 \over 2 H_z^2} \ ,
\end{eqnarray}
where $H_y$ and $H_z$ are given in Eq. (\ref{HYZEQ}), so that
\begin{eqnarray}
\sin (2 \theta) &=& \left( {2 \over 3J_1} \right)
{J_{yz} - C_{yz} - \sqrt 3 D_y \over \left[ 1 + {Z \over J_1} \right] } \ ,
\label{ZEQ} \end{eqnarray}
where
\begin{eqnarray}
Z &=& J_2 + {\Delta_J \over 4} - {\eta_J \over 12} - {\sqrt 3 D_z \over 3}
+ {C_{y-z} \over 3} \ .
\end{eqnarray}
Then
\begin{eqnarray}
\cos (2 \theta) = 1 - {2 \over 9J_1^2} 
(J_{yz} - C_{yz} - \sqrt 3 D_y )^2 \ .
\end{eqnarray}
Equation (\ref{TILDE}) gives
\begin{eqnarray}
\overline \Delta &\equiv & C_{xx} + \overline C_{yy}
= C_{xx} + C_{yy} - 2C_{zz} - 3 C_{yz} \sin (2 \theta)  \nonumber \\
&=& \Delta - 3C_{yz} \sin (2 \theta) \  .
\end{eqnarray}

Then, keeping only relevant anisotropy corrections, we find that
the constants in Eq. (\ref{NMATAEQ}) and (\ref{NMATBEQ}) are
\begin{eqnarray}
A_0 &=& 4 \overline E_z^{(1)} + \overline \Delta  \nonumber \\ &=&
\Delta + 2E_z^{(1)} + 2E_y^{(1)} + [2E_z^{(1)}-2E_y^{(1)}]
\cos (2 \theta) \nonumber \\ && \
- [4E_{yz}^{(1)}+3C_{yz}] \sin (2 \theta) \nonumber \\
&=& \Delta  + {3 \over 2} J_{xx} - {1 \over 2} J_{yy} - 
2 J_{zz} - \sqrt 3 D_z \nonumber \\ && \ 
+ \Biggl[ {3 \over 2} J_{xx} - {1 \over 2} J_{yy} + 2 J_{zz}
- \sqrt 3 D_z \Biggr]\nonumber \\ && \ \times  \Biggl[
1 - {2[J_{yz} - C_{yz} - \sqrt 3 D_y]^2 \over 9J_1^2} \Biggr] \nonumber \\
&& - {1 \over 3J_1} \Biggl[ 4\sqrt 3 D_y - 4J_{yz} + 6 C_{yz} \Biggr] \Biggl[
J_{yz} - C_{yz} - \sqrt 3 D_y \Biggr] \nonumber \\
&=& 2J + \Delta + {5 \over 6} \Delta_J + {3 \over 2} \eta_J 
- 2 \sqrt 3 D_z \nonumber \\ && \
+ {2  \over 3J_1} [J_{yz} - C_{yz} - \sqrt 3 D_y][J_{yz}
- \sqrt 3 D_y - 2 C_{yz}] \ ,  \nonumber \\
A_1 &=& - E_x^{(1)} - \overline E_y^{(1)}\nonumber \\
&=& - E_x^{(1)} - [E_z^{(1)} + E_y^{(1)}]/2\nonumber \\ && \
 - {1 \over 2}
[E_y^{(1)} - E_z^{(1)}] \cos (2 \theta) 
- E_{yz}^{(1)} \sin (2 \theta) \nonumber \\
&=&  {1 \over 4} J_{xx} - {3 \over 4} J_{yy} + {\sqrt 3 \over 2} D_z
- {3 \over 8} J_{xx}\nonumber \\ && \
 + {1 \over 8} J_{yy} + {1 \over 2} J_{zz}
+ {\sqrt 3 \over 4} D_z  \nonumber \\
&& \  + \Biggl[ {1 \over 2} J_{zz} + {3 \over 8} J_{xx} - {1 \over 8} J_{yy}
- {\sqrt 3 \over 4} D_z \Biggr] \nonumber \\ && \ \times
\Biggl[  1 - {2[J_{yz} - C_{yz} - \sqrt 3 D_y]^2 \over 9J_1^2} \Biggr]
\nonumber \\ && \ - {1 \over 3J_1} [\sqrt 3 D_y - J_{yz}]
[J_{yz} - C_{yz} - \sqrt 3 D_y] \nonumber \\ 
&=& {1 \over 2} J_1 - {3 \over 8} \eta_J + {11 \over 24} \Delta_J
+ {\sqrt 3 \over 2} D_z  \nonumber \\ && \
+ {1 \over 6J_1} \Biggl[ J_{yz} - C_{yz} - \sqrt 3 D_y \Biggr] \Biggl[
- \sqrt 3 D_y + J_{yz} + C_{yz} \Biggr] \ , \nonumber \\
B_0 &=& C_{xx}- \overline C_{yy} \ , \nonumber \\ 
&=& C_{xx} - {1 \over 2} \Biggl[ C_{yy} + C_{zz} \Biggr]
+ {1 \over 2} \Biggl[ C_{zz}-C_{yy} \Biggr] \cos (2 \theta)
\nonumber \\ && \ + C_{yz} \sin (2\theta) \nonumber \\
&=& C_{xx}-C_{yy} + {2C_{yz} \over 3J_1} [J_{yz} - C_{yz} - \sqrt 3 D_y] \ , 
\nonumber \\
B_1 &=& - E_x^{(1)} + \overline E_y^{(1)} \nonumber \\ &=&
-E_x^{(1)} + {1 \over 2} \Biggl[ E_y^{(1)} + E_z^{(1)} \Biggr] 
+ E_{yz}^{(1)} \sin (2 \theta) \nonumber \\ && \ +
{1 \over 2} \Biggl[ E_y^{(1)}-E_z^{(1)} \Biggr] \cos (2 \theta) 
\nonumber \\
&=& {5 \over 8} J_{xx} - {7 \over 8} J_{yy} - {1 \over 2} J_{zz}
+ {\sqrt 3 \over 4} D_z \nonumber \\ && \ 
- {1 \over 2} \Biggl[ J_{zz} + {3 \over 4} J_{xx} - {1 \over 4} J_{yy}
- { \sqrt 3 \over 2} D_z \Biggr]
\nonumber \\ && \ \times
\Biggl[  1 - {2[J_{yz} - C_{yz} - \sqrt 3 D_y]^2 \over 9J_1^2} \Biggr]
\nonumber \\ && \ + {1 \over 3J_1} [ \sqrt 3 D_y - J_{yz}][
J_{yz} - C_{yz} - \sqrt 3 D_y] \nonumber \\ 
&=& - {3 \over 2} J_1 + {5 \over 8} \eta_J + {1 \over 8} \Delta_J
+ {\sqrt 3 \over 2} D_z\nonumber \\ && \ 
 + {[J_{yz}-C_{yz} - \sqrt 3 D_y] \over 6J}
[\sqrt 3 D_y - J_{yz} - C_{yz}] \ , \nonumber \\
\alpha^{(1)} &=& 2 \overline d_z^{(1)}
= 2d_z^{(1)} \cos \theta + 2d_y^{(1)} \sin \theta \nonumber \\
&=& 2d_z^{(1)} + d_y^{(1)} \sin (2 \theta) \nonumber \\
&=& -(D_y + \sqrt 3 J_{yz}) \cos \theta \nonumber \\ &+&
\left(- \sqrt 3 J_1 - D_z + {\sqrt 3 \Delta_J \over 12}
- { \sqrt 3 \eta_J \over 4} \right) \sin \theta \ .
\end{eqnarray}
But using Eq. (\ref{ZEQ}) this gives
\begin{eqnarray}
\alpha^{(1)} &=& - D_y - \sqrt 3 J_{yz} + {J_{yz} - C_{yz} - \sqrt 3 D_y 
\over 3J_1} \nonumber \\ && \ \times \left[ 1 - {Z \over J_1} \right]
\left[ - \sqrt 3 J_1 - D_z + {\sqrt 3 \Delta_J \over 12} - {\sqrt 3 \eta_J
\over 4} \right] \nonumber \\ &=&
- {4 \sqrt 3 \over 3} J_{yz} + {C_{yz} \over \sqrt 3}
+ {J_{yz} - C_{yz} - \sqrt 3 D_y \over 3 \sqrt 3 J_1} \nonumber \\ && \
\times \left[ \Delta_J - \eta_J - 2 \sqrt 3 D_z + C_{y-x} + 3J_2 \right] \ .
\label{ALPHAEQ} \end{eqnarray}
Also
\begin{eqnarray}
\alpha^{(2)} &=& 2 d_z^{(2)} \cos \theta + 2 d_y^{(2)} \sin \theta \nonumber \\
&=& (- \sqrt 3 J_2) \sin \theta \nonumber \\ &=&
- {\sqrt 3 J_2 \over 3 J_1} \left[ J_{yz} - C_{yz} - \sqrt 3 D_y \right] \ .
\label{ALPHAEQ2} \end{eqnarray}

\section{DYNAMICAL STRUCTURE FACTOR}

We first define the Green's function
\begin{eqnarray}
\langle \langle A;B \rangle \rangle_\omega &=&
\sum_n p_n \Biggl[ { \langle n | A | m \rangle \langle m | B | n \rangle
\over \omega - E_m + E_n } \nonumber \\ && \
- { \langle n | B | m \rangle \langle m | A | n \rangle
\over \omega + E_m - E_n }  \Biggr] \ .
\end{eqnarray}
Then we will need the frequency and wave vector dependent susceptibility
given by
\begin{eqnarray}
\chi^{\alpha \beta} ( \omega, {\bf q}) &=&
\langle \langle {\cal S}^\alpha ({\bf q});
{\cal S}^\beta (-{\bf q}) \rangle \rangle_\omega
\end{eqnarray}
in terms of which the dynamical structure factor is
\begin{eqnarray}
{\cal S} (\omega , {\bf q}) &=& {n(\omega) \over \pi} \sum_{\alpha \beta}
\Im \chi^{\alpha \beta}(\omega- i 0^+, {\bf q}) [ \delta_{\alpha , \beta}
- \hat q_\alpha \hat q_\beta] \ ,
\end{eqnarray}
where $n(\omega) = [ e^{\beta \omega} -1]^{-1}$.  We have that
\begin{eqnarray}
&& {n(\omega) \over \pi} \Im \chi^{\alpha \beta} ( \omega - i0^+, {\bf q}) 
\nonumber \\ && \ =
\sum_{m,n} p_n \Biggl[ \delta(\omega-E_m + E_n) \langle n|{\cal S}^{\alpha}
({\bf q}) |m \rangle \langle m| {\cal S}^{\beta} (-{\bf q})|n \rangle
\nonumber \\ && \ \
- \delta(\omega + E_m - E_n) \langle n|{\cal S}^{\beta}
(-{\bf q}) |m \rangle \langle m| {\cal S}^{\alpha} ({\bf q})|n \rangle
\Biggr] n(\omega ) \nonumber \\
&=& \sum_{m,n} \Biggl\{ p_n \delta(\omega-E_m + E_n) \langle n|{\cal S}^{\alpha}
({\bf q}) |m \rangle \langle m| {\cal S}^{\beta} (-{\bf q})|n \rangle
\nonumber \\ && \ \
- p_m \delta(\omega + E_n - E_m) \langle m|{\cal S}^{\beta}
(-{\bf q}) |n \rangle \langle n| {\cal S}^{\alpha} ({\bf q})|m \rangle
\Biggr\} n(\omega) \nonumber \\
&=& \sum_{m,n} p_n \delta(\omega-E_m + E_n) [1 - e^{-\beta(E_m-E_n)}]
n(\omega ) \nonumber \\ && \ \ \times
\langle n|{\cal S}^{\alpha} ({\bf q}) |m \rangle
\langle m| {\cal S}^{\beta} (-{\bf q})|n \rangle \nonumber \\
&=& \sum_{m,n} p_n \delta(\omega-E_m + E_n) {1 - e^{-\beta(E_m-E_n)}]
\over e^{\beta(E_m-E_n)} -1 } \nonumber \\ && \ \ \times
\langle n|{\cal S}^{\alpha} ({\bf q}) |m \rangle
\langle m| {\cal S}^{\beta} (-{\bf q})|n \rangle \nonumber \\
&=& \sum_{m,n} p_m \delta(\omega-E_m + E_n)
\langle n|{\cal S}^{\alpha} ({\bf q}) |m \rangle
\langle m| {\cal S}^{\beta} (-{\bf q})|n \rangle \ .
\end{eqnarray}

We write
\begin{eqnarray}
{\cal S}^\alpha ({\bf q}) &=&
{1 \over \sqrt N} \sum_i e^{i {\bf q} \cdot {\bf r}_i}
{\cal S}^\alpha (i) \nonumber \\ &=& {1 \over \sqrt N} \sum_{\tauv , {\bf R}}
e^{i {\bf q} \cdot (\tauv + {\bf R})}
{\cal S}^\alpha ({\bf R}, \tauv ) \nonumber \\ &=&
{1 \over \sqrt N} \sum_{\tauv , {\bf R}, \rho}
e^{i {\bf q} \cdot (\tauv + {\bf R})}
{\cal R}_{\alpha \rho} ( \tau) S^\rho_\ell ({\bf R}, \tauv ) \ ,
\end{eqnarray}
where the subscript $\ell$ indicates a component with respect to the
local canted axes. Then, to linear order in the Bose operators we have
\begin{eqnarray}
{\cal S}^\alpha ({\bf q}) &=&
{1 \over \sqrt N} \sum_{\tauv , {\bf R} }
e^{i {\bf q} \cdot (\tauv + {\bf R})} \nonumber \\ && \ \times \Biggl[
{\cal R}_{\alpha x} ( \tau) S^x_\ell ({\bf R}, \tauv )
+ {\cal R}_{\alpha y} ( \tau) S^y_\ell ({\bf R}, \tauv ) \Biggr]
\nonumber \\ &=& \sqrt {S \over 2 N} \sum_{\tauv , {\bf R} }
e^{i {\bf q} \cdot (\tauv + {\bf R})} \nonumber \\ && \ \times \Biggl(
{\cal R}_{\alpha x} ( \tau) [a({\bf R}, \tauv )+ a^\dagger({\bf R} , \tauv) ]
\nonumber \\ && \ \ 
-i {\cal R}_{\alpha y} ( \tau) [a({\bf R}, \tauv ) - a^\dagger({\bf R} , \tauv)
] \Biggr) \ .
\end{eqnarray}
Now set
\begin{eqnarray}
\Gamma_\alpha (\tau) &\equiv& {\cal R}_{\alpha x}(\tau)
+ i {\cal R}_{\alpha y} (\tau) \ .
\end{eqnarray}
Then
\begin{scriptsize}
\begin{eqnarray}
{\cal S}^\alpha ({\bf q}) &=& \sqrt {S \over 2N} \sum_{{\bf R} \tauv}
e^{i {\bf q} \cdot ({\bf R}+\tauv)}
\left[ \Gamma_\alpha (\tau) a^\dagger ({\bf R} \tau) +
\Gamma_\alpha (\tau)^* a({\bf R} \tau) \right] \nonumber \\ &=&
\sqrt{S \over 2} \sum_\tauv \Biggl[ 
\Gamma_\alpha (\tau) a^\dagger_\tau ({\bf q}) +
\Gamma_\alpha (\tau)^* a_\tau (-{\bf q}) \Biggr] \nonumber \\ &=&
\sqrt{S \over 2} \sum_{\tauv \mu} \Biggl[
\Gamma_\alpha (\tau) \Biggl( c^{(\mu)}_\tau ({\bf q})^* X_\mu^\dagger ({\bf q})
- d_\tau^{(\mu)} ({\bf q}) X_\mu(-{\bf q})\Biggr) \nonumber \\ && \ \
+ \Gamma_\alpha (\tau)^* \Biggl( c_\tau^{(\mu)}(-{\bf q}) X_\mu(-{\bf q})
- d_\tau^{(\mu)} (-{\bf q})^* X_\mu^\dagger ({\bf q})\Biggr) \Biggr]
\nonumber \\ &=& \sqrt{S \over 2} \sum_{\tauv \mu} \Biggl[ \Biggl(
\Gamma_\alpha (\tau) c_\tau^{(\mu)}({\bf q})^*
- \Gamma_\alpha(\tau)^* d_\tau^{(\mu)}
({\bf q})^* \Biggr) X_\mu^\dagger ({\bf q}) \nonumber \\ && \ \
+ \Biggl( \Gamma_\alpha (\tau)^* c_\tau^{(\mu)}({\bf q})
- \Gamma_\alpha(\tau) d_\tau^{(\mu)} ({\bf q}) \Biggr)
X_\mu(-{\bf q}) \Biggr] \ .
\end{eqnarray}
\end{scriptsize}
For general reciprocal lattice vector one has
\begin{small}
\begin{eqnarray}
&& {\cal S}^\alpha ({\bf q}+{\bf G}) = \sqrt{S/2} \nonumber \\ && \ \times
\sum_{\tauv \mu} \Biggl[ \Biggl(
\Gamma_\alpha (\tau) c_\tau^{(\mu)}({\bf q})^*
- \Gamma_\alpha(\tau)^* d_\tau^{(\mu)} ({\bf q})^* \Biggr)
X_\mu^\dagger ({\bf q}) \nonumber \\ && \ \
+ \Biggl( \Gamma_\alpha (\tau)^* c_\tau^{(\mu)}({\bf q})
- \Gamma_\alpha(\tau) d_\tau^{(\mu)} ({\bf q}) \Biggr) X_\mu(-{\bf q}) \Biggr]
e^{i {\bf G} \cdot \tauv}
\end{eqnarray}
\end{small}
This result indicates that the scattering intensity will oscillate
when $e^{i{\bf G} \cdot \tauv}$ changes sign as one goes from one
Brillouin zone to the next.

\end{document}